\documentclass{emulateapj}

\usepackage{psfig}
\usepackage{natbib}
\usepackage{txfonts}
\usepackage{lscape}
\usepackage{longtable}

\begin{document}
\def\lsun{L_{\sun}}
\def\msun{M_{\sun}}
\def\simle{{\mathop{\stackrel{\sim}{\scriptstyle <}}\nolimits}}
\def\simgr{{\mathop{\stackrel{\sim}{\scriptstyle >}}\nolimits}}
\def\lesim{{\mathop{\stackrel{\scriptstyle <}{\sim}}\nolimits}}
\shorttitle{VLA observations of NH$_3$ in IRDCs}
\shortauthors{Ragan, Bergin, \& Wilner}

\title{Very Large Array Observations of Ammonia in Infrared-Dark Clouds I: Column Density and Temperature Structure}

\author{Sarah E. Ragan\altaffilmark{1,2}, Edwin A. Bergin\altaffilmark{1}, David Wilner\altaffilmark{3}}

\altaffiltext{1}{Department of Astronomy, University of Michigan, 830 Dennison Building, 500 Church Street, Ann Arbor, MI, 48109 USA}
\altaffiltext{2}{Max Planck Institute for Astronomy, K\"{o}nigstuhl 17, 69117 Heidelberg, Germany}
\altaffiltext{3}{Smithsonian Astrophysical Observatory, Cambridge, MA, USA}

\begin{abstract}

We present Very Large Array observations of NH$_3$ (1,1) and (2,2) in a sample of six infrared-dark clouds (IRDCs) with distances from 2 to 5~kpc.   We find that ammonia serves as an excellent tracer of dense gas in IRDCs, showing no evidence of depletion, and the average abundance in these clouds is 8.1 $\times$ 10$^{-7}$. Our sample consists of four IRDCs with 24~$\mu$m embedded protostars and two that appear starless.  We calculate the kinetic temperature of the gas in IRDCs and find no significant difference between starless and star-forming IRDCs.  We find that the bulk of the gas is between 8 and 13~K, indicating that any embedded or nearby stars or clusters do not affect the gas temperature dramatically.  
Though IRDCs have temperatures and volume densities on par with local star formation regions of lower mass, they consist of much more mass which induces very high internal pressures.  In order for IRDCs to survive as coherent structures, the internal pressure must be balanced by a confining pressure provided by the high concentration of molecular clouds in the spiral arm in which they reside.   The high molecular concentration and pressure is roughly consistent with gas dynamics of a bar galaxy.

\end{abstract}

\section{Introduction}
\label{intro}

Tracing the evolution of the Galaxy relies heavily on what is known of star formation.  Here, in the outskirts of the Galactic disk near the Sun, however, only a part of the picture is represented.  Low-mass star formation dominates the local neighborhood, while the massive stars that dominate the energetics in the Galaxy, lie much further away.  If we are to follow in the footsteps of the successful studies of nearby regions of star formation, it is essential to isolate and characterize objects in the evolutionary stages leading the formation of a massive star and cluster.  This poses a daunting challenge because, as we've learned from low mass studies, such objects in early stages are cold, dark, and obscured by heavily extincted envelopes, and these conditions are only more extreme in the case of more massive star formation.  Thus, we must turn from the optical and near infrared studies and look longward in wavelength where dust and molecules reveal the nature of the early stages of massive star formation.

Over the past decade, infrared-dark clouds (IRDCs) have become the front-runners for the precursors to massive stars and clusters \citep[e.g.][]{Carey_submmIRDC,rathborne2006}.  First, their temperatures, densities, and masses (T $<$ 20~K, $n >$ 10$^5$~cm$^{-3}$, M $\sim$ 10$^2$ - 10$^3~\msun$) suit the conditions required for massive stars to form \citep[e.g.][]{ragan_msxsurv,Simon2006b}.  Furthermore, their substructure \citep{ButlerTan2009, Hennemann2009}, including their clump mass function \citep{Ragan_spitzer,PerettoFuller2009}, mass-radius relationship \citep{Kauffmann_masssize1} show that IRDCs are undergoing fragmentation.  Finally, observations of pre- and protostellar cores \citep{A&ASpecialIssue-Henning} within IRDC filaments, young stellar objects on the IRDC surface \citep{Ragan_spitzer}, and embedded massive stars in the densest regions \citep[e.g.][]{Rathborne_2007_protostars, Beuther_protostars_IRDC} show that IRDCs are actively forming stars, some of which are massive. 

This wealth of new insight into the nature of IRDCs has resulted in large part from the advent of Galactic Plane surveys in the mid-infrared (MSX, ISO, Spitzer), where IRDCs appear as absorbing structures against the background provided primarily by polycyclic aeromatic hydrocarbon emission.  Follow up studies using  single-dish observations of dust continuum emission have also been used as a benchmark for the gross properties of IRDCs, but such surveys suffer from insufficient angular resolution to probe the ``clump'' and ``core'' scales seen with Spitzer.  Only a handful of interferometric studies have been done \citep[e.g.][]{Hennemann2009,BeutherHenning2009}, but our capability to conduct such surveys will markedly increase with ALMA.  In the meantime, there is much to be learned about IRDCs from the study of molecular line emission.  

The ammonia molecule (NH$_3$) has been widely used to characterize the physical conditions in nearby molecular clouds \citep[e.g.][]{Wiseman1998, Jijina1999, Li_oriontemp, Rosolowsky_perseus, Friesen2009}.  A symmetric top molecule, NH$_3$ is a valuable tool because of its sensitivity to a broad range of excitation conditions and it probes the very densest regions of molecular cloud cores. In particular, the lower metastable states of the vibrational inversion transitions, ($J,K$) =  (1,1) and (2,2), are excited at the characteristically low ($<$ 20~K) temperatures of molecular clouds and IRDCs.  The splitting of the inversion transitions into hyperfine structure, due to the interaction between the N nucleus' electric quadrupole moment and the electric field of the electrons, allows for the straightforward calculation of optical depth, column density, and gas temperature \citep[see][for derivation details]{HoTownes}.  In these nearby clouds, ammonia is found to preferentially trace the densest material with temperatures between 10 and 20~K.  It is therefore of interest to test whether this correlation holds in the IRDC environment where clusters and massive stars are forming.

Beyond local molecular clouds, \citet{Pillai_ammonia} mapped ammonia in nine IRDCs using a single dish telescope with resolution $\sim$ 40\arcsec. They noted that some IRDCs show spatial temperature gradients on large scales.  However, these observations do not resolve the dense cores and clumps that exist on 10~$''$ scales \citep{Ragan_spitzer}.  Furthermore, {\em Spitzer} has revealed the presence of young stellar populations on the surface of and within IRDCs, and the possible resulting temperature variation on these small spatial scales has not been examined systematically.    

We present observations towards a sample of six infrared-dark clouds from the \citep{Ragan_spitzer} sample of the NH$_3$ ($J,K$) = (1,1) and (2,2) inversion transitions using combined data from the Very Large Array (VLA) and Green Bank Telescope (GBT).  We relate the distribution of ammonia to the column density distribution in IRDCs, as well as the absence and presence of embedded star formation, from our {\em Spitzer} study \citep{Ragan_spitzer}.  
This sample exhibits a mixture of apparently starless IRDC clumps and regions of active star formation, and we thus relate the star formation properties to signatures seen in the lines.
Using the unique properties of the ammonia transitions as a gas thermometer \citep{Walmsley1983}, we map the kinetic temperature of the gas at angular resolution between 4 and 8$\arcsec$, corresponding to 0.08 and 0.16~pc spatial scales at 4~kpc.  This study addresses the changes in gas temperature on the spatial scales relevant to cloud fragmentation.  The kinematic characteristics of these objects are not analyzed here, but will be addressed in a future paper.

\begin{table*} \centering
\caption[Observation Summary]{Target Summary. IRDC positions, distance, rms and beamsizes of the combined dataset. \label{tab:obssumm}}
\begin{tabular}{lcccccc}
\hline
IRDC & $\alpha$ & $\delta$ & D & v$_{lsr}$ &  rms & beam size \\
& (J2000) & (J2000) & (kpc) & (km s$^{-1}$) &  (mJy) & ($\arcsec \times \arcsec$) \\
\hline
G005.85$-$0.23 &
17:59:53 &
$-$24:00:10 & 3.14 &
17.2 & 
2.8 & 7.7 $\times$ 6.8 \\

G009.28$-$0.15 &
18:06:54 &
$-$20:58:51 & 4.48 &
41.4 & 
4.8 & 8.3 $\times$ 6.4 \\

G009.86$-$0.04 &
18:07:40 &
$-$20:25:25 & 2.36 &
18.1 & 
4.3 & 8.1 $\times$ 6.3 \\

G023.37$-$0.29 & 
18:34:51 &
$-$08:38:58 & 4.70 &
78.5 &
2.5 & 5.7 $\times$ 3.7 \\

G024.05$-$0.22 &  
18:35:52 & 
$-$08:00:38 & 4.82 &
81.4 &
4.3 & 8.2 $\times$ 7.0 \\

G034.74$-$0.12 &
18:55:14 & 
$+$01:33:42 & 4.86 &
79.1 &
6.8 & 8.1 $\times$ 7.0 \\

\hline
\end{tabular}
\end{table*}

\section{Observations \& Data Reduction}
\label{obs}

We select a subsample of six IRDCs from the \citet{Ragan_spitzer} survey.  This sample is biased toward the most compact and opaque objects, in hopes that this increases the possibility of isolating massive starless (or pre-stellar) cores and clumps.  The observation details can be found in Table~\ref{tab:obssumm}.  We obtained both single dish and array data of the NH$_3$ (1,1) and (2,2) lines.  

\subsection{Green Bank Telescope}

We acquired single-dish observations of NH$_3$ ($J,K$) = (1,1) and (2,2) inversion lines using the Robert C. Byrd Green Bank Telescope (GBT) from 6 to 15 September 2006.  The rest frequencies of the (1,1) and (2,2) lines are 23.6944955 and 23.7226336~GHz \citep{HoTownes}, respectively.  At these frequencies, the beam FWHM was approximately 32$''$.  These upper K-band observations were made in frequency-switching mode.  The GBT spectrometer back end was configured to simultaneously observe the two transitions in separate spectral windows, using $\sim$32,768 spectral channels in each IF that were 50~MHz wide.  The spectral resolution for the (1,1) line was 0.03~km~s$^{-1}$, or 1.6~kHz.  

We mapped the regions using a single-pointing grid with Nyquist sampling, the parameters of which are summarized in Table~\ref{tab:obssumm}.  Integration times were 2-3 minutes per point, depending on the strength of the line, elevation of the source, and thus the time required to obtain sufficient signal-to-noise.  Pointing corrections were done toward various calibrators depending on the time of observation (1733-1304, 1814-1853, 1822-0938, 1831-0949, and 1851+0035) every 45 to 75 minutes, weather-depending, which resulted in corrections of a few arcseconds, typically.  System temperatures were between 60 and 100~K, and the elevation of the sources ranged between 20 and 40 throughout the observations.

Data were reduced and calibrated using GBTIDL\footnote{GBTIDL is an interactive package for the reduction and analysis of spectral line data taken with the GBT. http://gbtidl.nrao.edu/}.  The frequency-switched observations were reduced with the {\tt getfs} routine, which retrieves, calibrates, and plots the spectrum.  We then applied {\tt hanning} smoothing and then {\tt boxcar} smoothing over 50 channel bins.  Five of the spectral components of the NH$_3$ signature were always spectrally resolved, and we performed gaussian fitting on each hyperfine component.  The data were then put into a FITS data cube using a homemade IDL-based script.

\subsection{Very Large Array}

Observations of the NH$_3$ (1,1) and (2,2) inversion transitions were undertaken of the sample in the compact D configuration of the Very Large Array (VLA) over the course of three tracks in 2007 April.    At this time, the array consisted of a hybrid of VLA and the updated EVLA receivers, and Doppler tracking was unavailable.  The observations were conducted in fixed-frequency mode.  Bandpass and calibration were done with observations of 1331$+$305 and 2253$+$161.  Phase calibration was done with observations of 1833-210, 1832-105, and 1851$+$005 every $\sim$12-15 minutes, and pointing was also performed on these sources every $\sim$ hour.  The data were flagged and calibrated using NRAO Astronomical Image Processing System (AIPS\footnote{http://www.aips.nrao.edu/}) according to the method described in the AIPS cookbook, taking into account corrections for the absence of Doppler tracking.  

\begin{figure}
\begin{center}
\includegraphics[scale=0.3,angle=90]{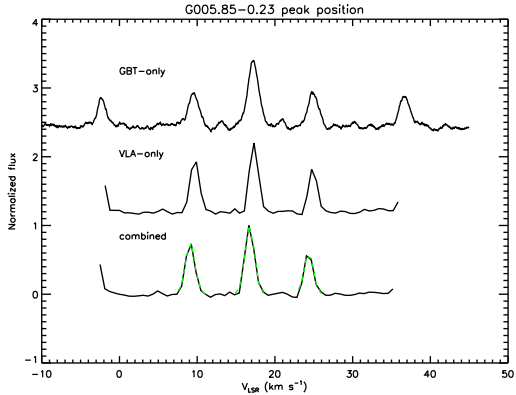}
\end{center}
\caption{Example spectrum from the central position of G005.85-0.23 from the GBT, VLA, and the combined data set. Overplotted in the dashed green line on the combined spectrum is our fit to the central three components of the NH$_3$(1,1) signature. \label{fig:spectrum}}
\end{figure}

These K-band observations were made using a four IF correlator backend configured with 3.125~MHz bandwidth containing 64 channels, yielding a spectral resolution of 48.8~kHz, or 0.6~km s$^{-1}$.  This setup was selected as to ensure detection of relatively weak lines in IRDCs, both the (1,1) and (2,2) simultaneously, with enough spectral resolution to resolve the lines.  Unlike the GBT setup, this configuration only fits the central three of the five main components of the ammonia signature (see Figure~\ref{fig:spectrum}). We discuss the effects of this in Section 3.  Table~\ref{tab:obssumm} gives the target, pointing and sensitivity information for each IRDC.  

\subsection{Single-dish and Interferometer Data Combination}

The detectable size of structures has an upper limit when using an interferometer, which is set by the shortest spacing between two antennas in the array.  Any flux on scales larger than this upper limit will be missing from the final image, a phenomenon known as the short-spacing problem.  Because IRDCs are complex structures, with emission on all scales, it is necessary to recover any large scale emission by combining the interferometer data with single-dish observations at the same frequency.  The 100-meter Green Bank Telescope offers the perfect complement to our D array observations with the Very Large Array.  

Following \citet{Friesen2009}, the GBT were first scaled and regridded to match the resolution of the VLA data.  Using the MIRIAD task {\tt IMMERGE}, the deconvolved interferometer images were combined in the Fourier domain with the GBT images.  {\tt IMMERGE} uses a tapering function which weights short spacings higher than long spacings.  Using an annulus ranging from 30 to 70 meters, where the high and low resolution images overlap, we ensure accurate determination of the flux calibration factor.  The combined data sets were binned in velocity to 0.6~km s$^{-1}$, which reduced the level of random errors.
This method was applied to the (1,1) and (2,2) datasets independently, yielding images of the same spatial and velocity resolution as the VLA image.  Table~\ref{tab:obssumm} gives the final beam dimensions, as well as other source parameters, for each object.

\begin{figure*}
\hbox{
\vspace{1.0cm}
\psfig{figure=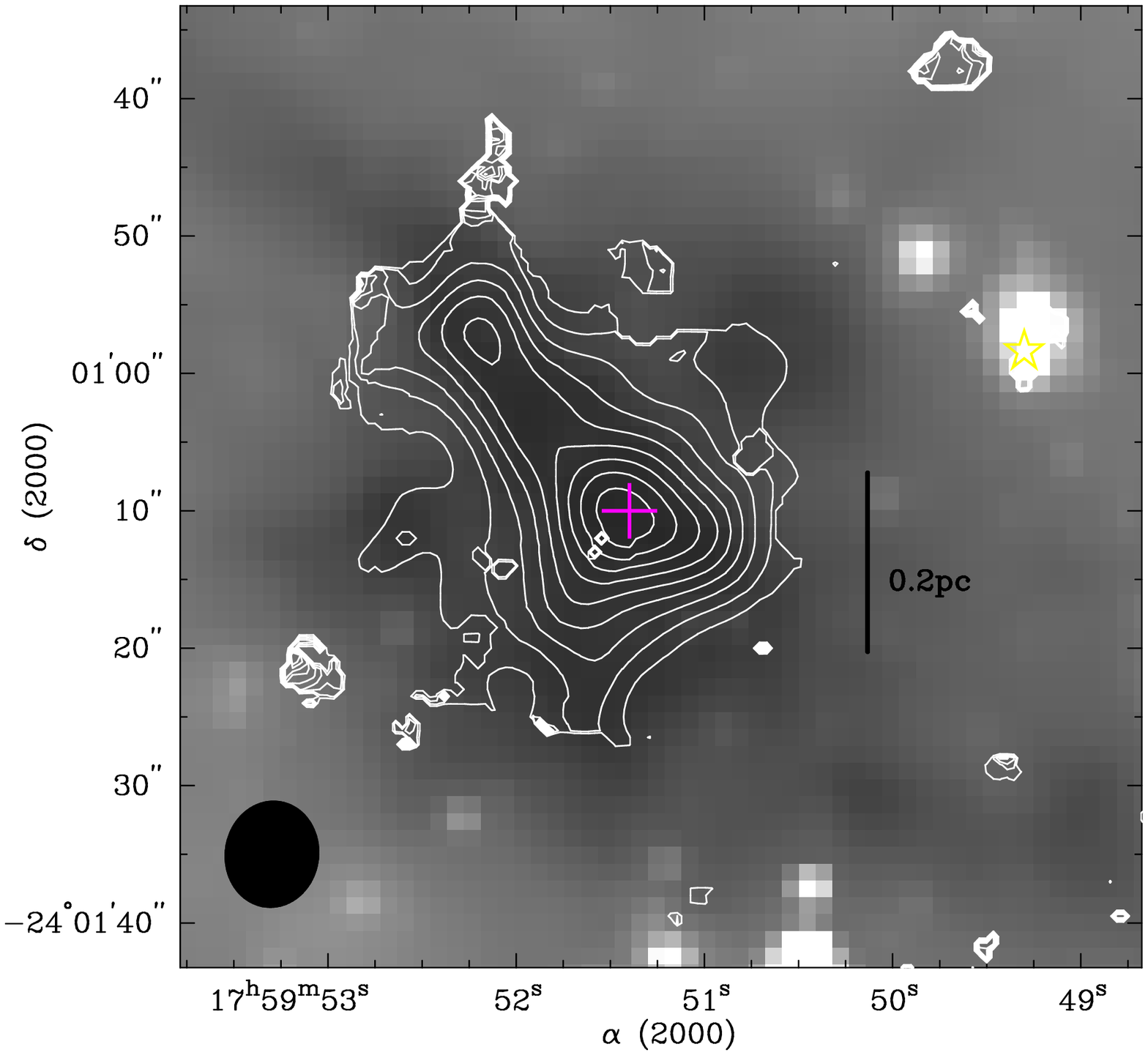,height=5.0cm,angle=270}
\hspace{0.3cm}
\psfig{figure=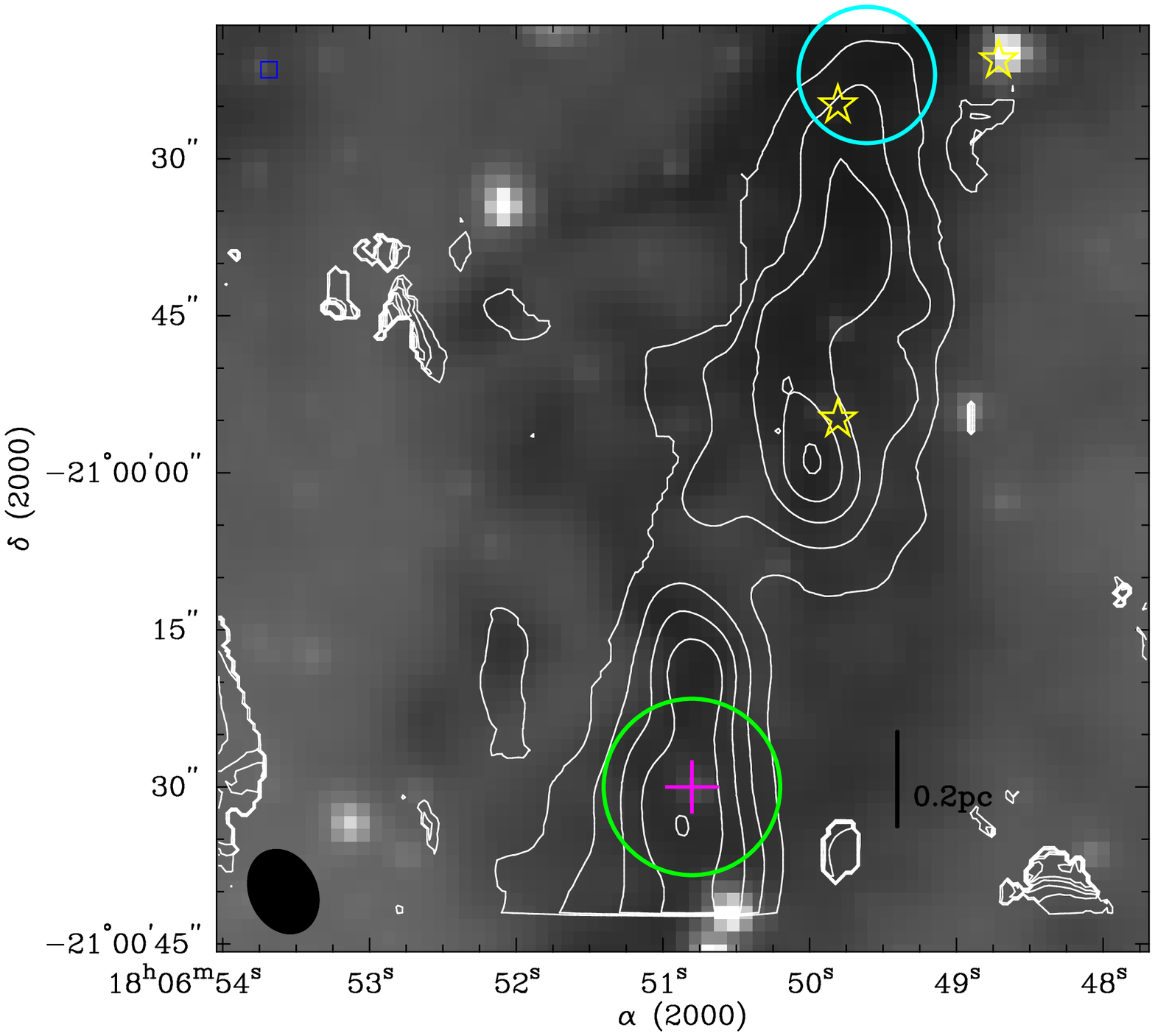,height=5.0cm,angle=270}
\hspace{0cm}
\psfig{figure=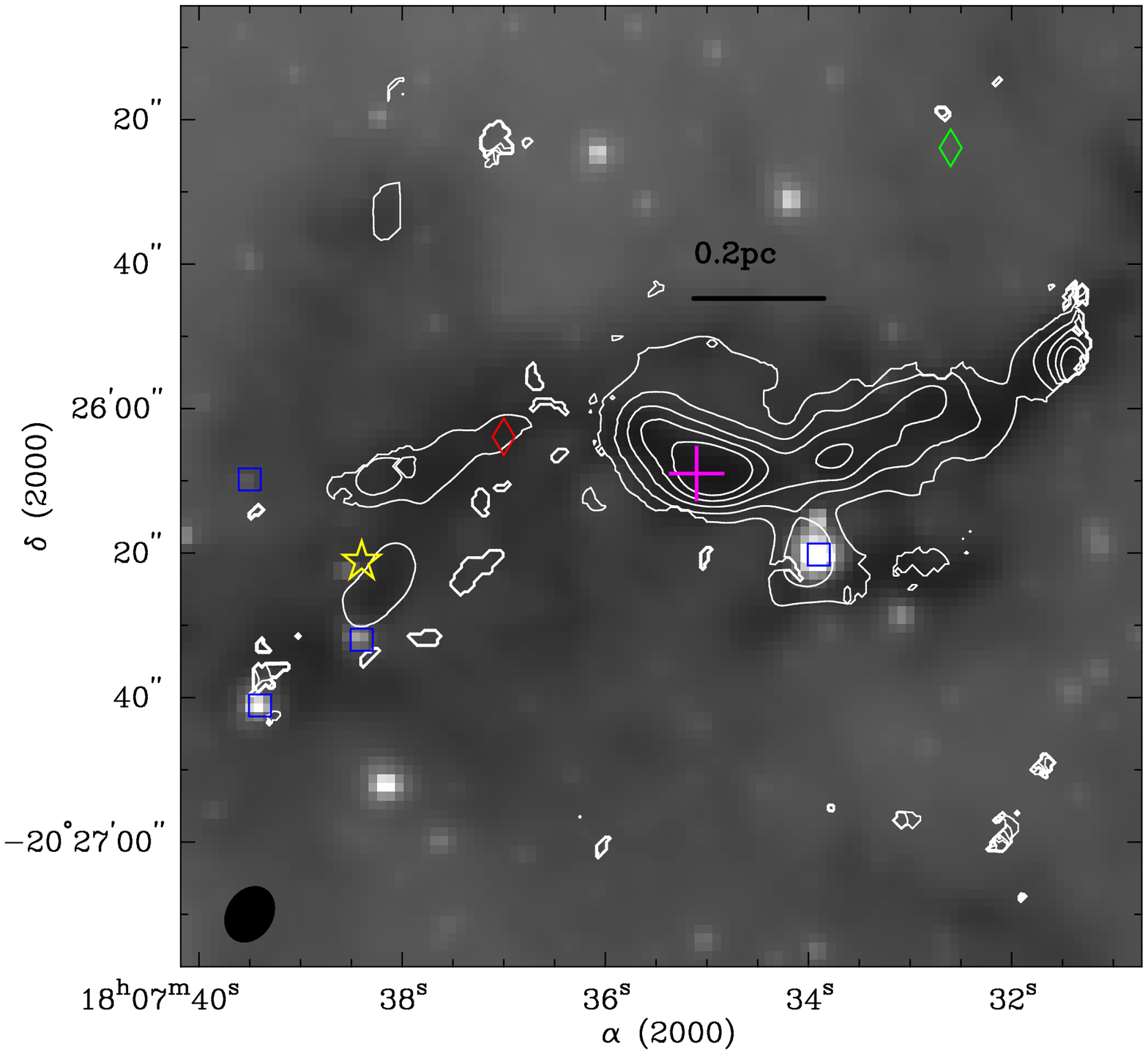,height=5.0cm,angle=270}
}
\vspace{-1.0cm}
\hbox{
\vspace{1.0cm}
\psfig{figure=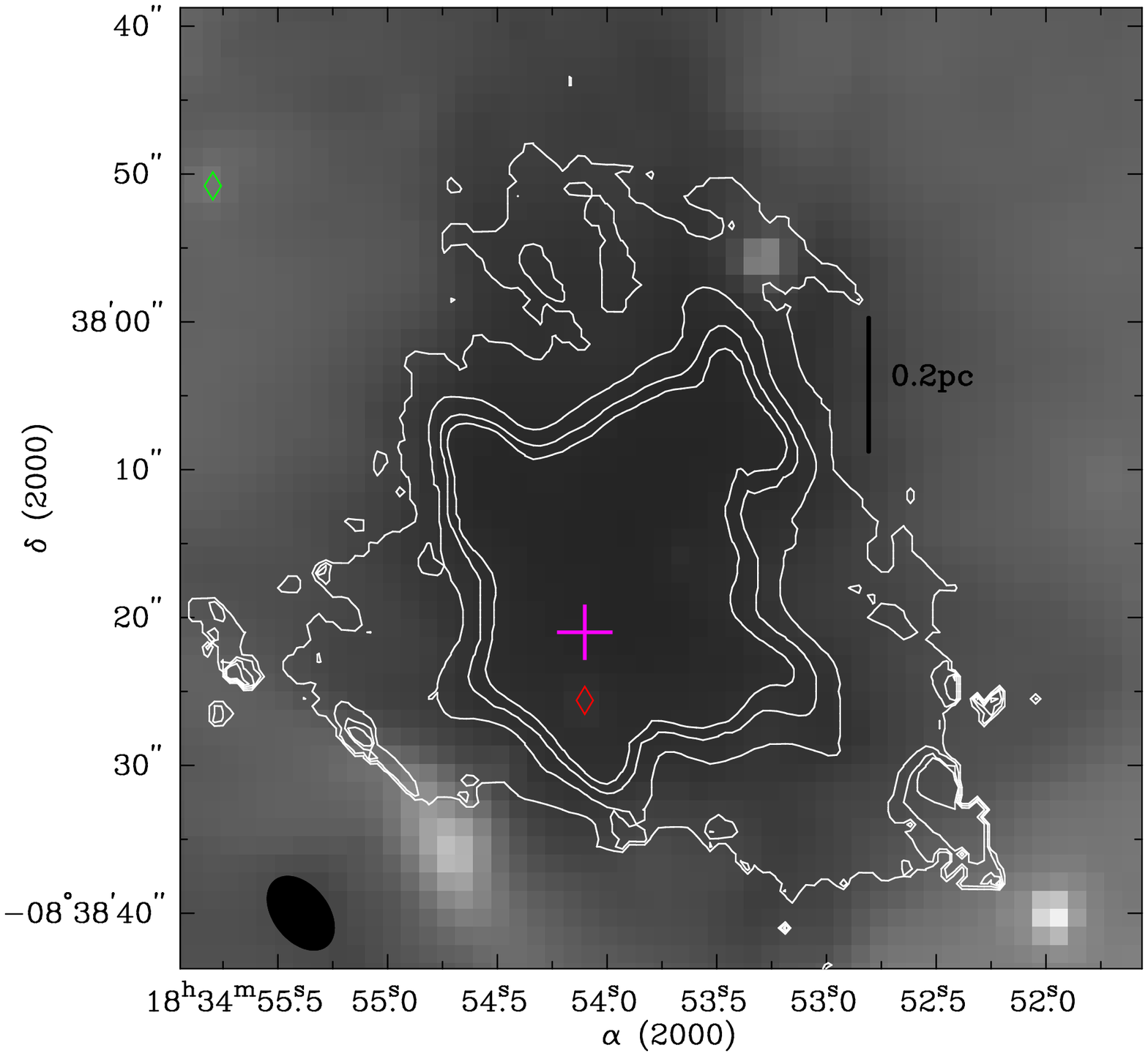,height=5.0cm,angle=270}
\hspace{0.5cm}
\psfig{figure=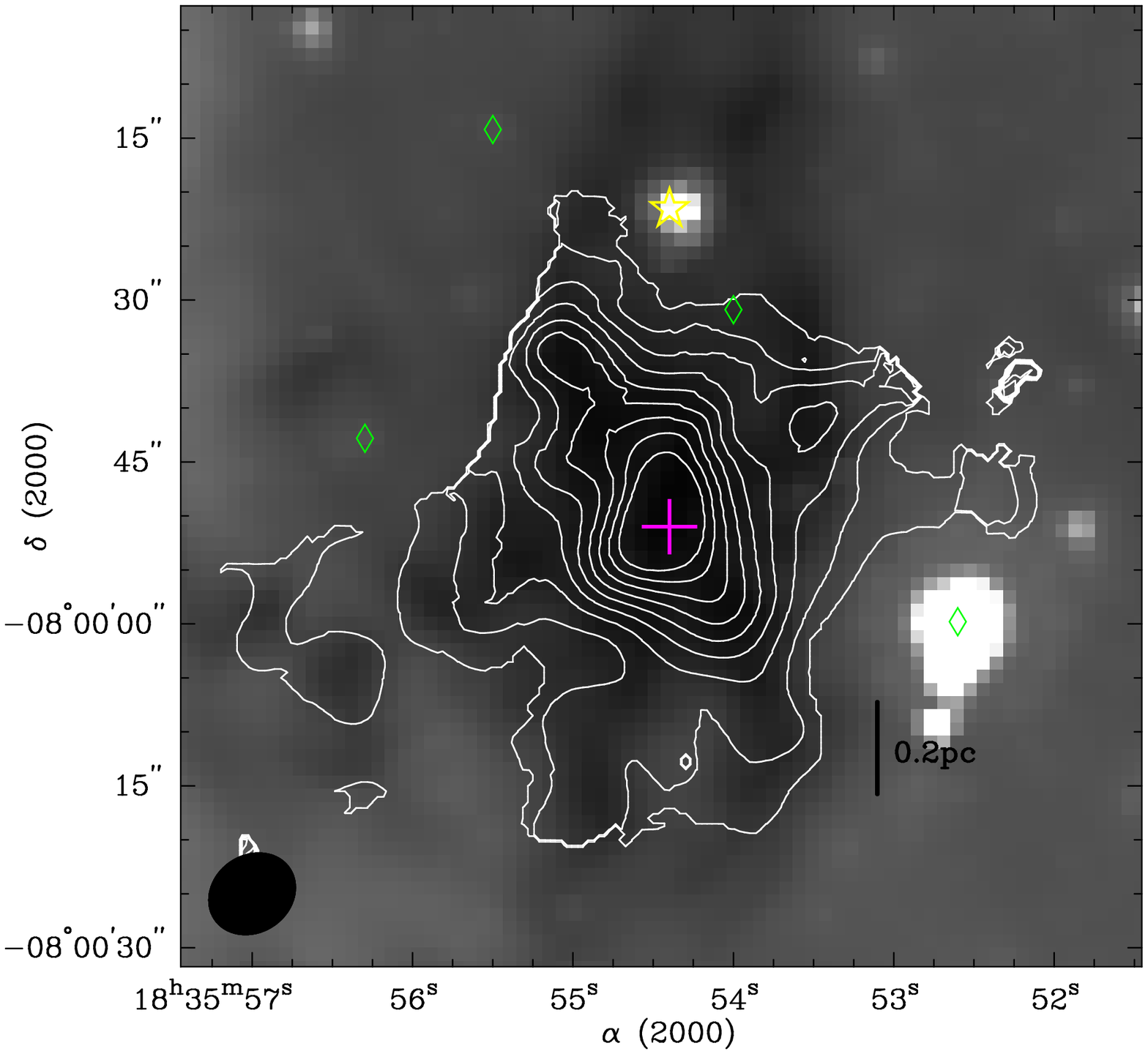,height=5.0cm,angle=270}
\hspace{0.3cm}
\psfig{figure=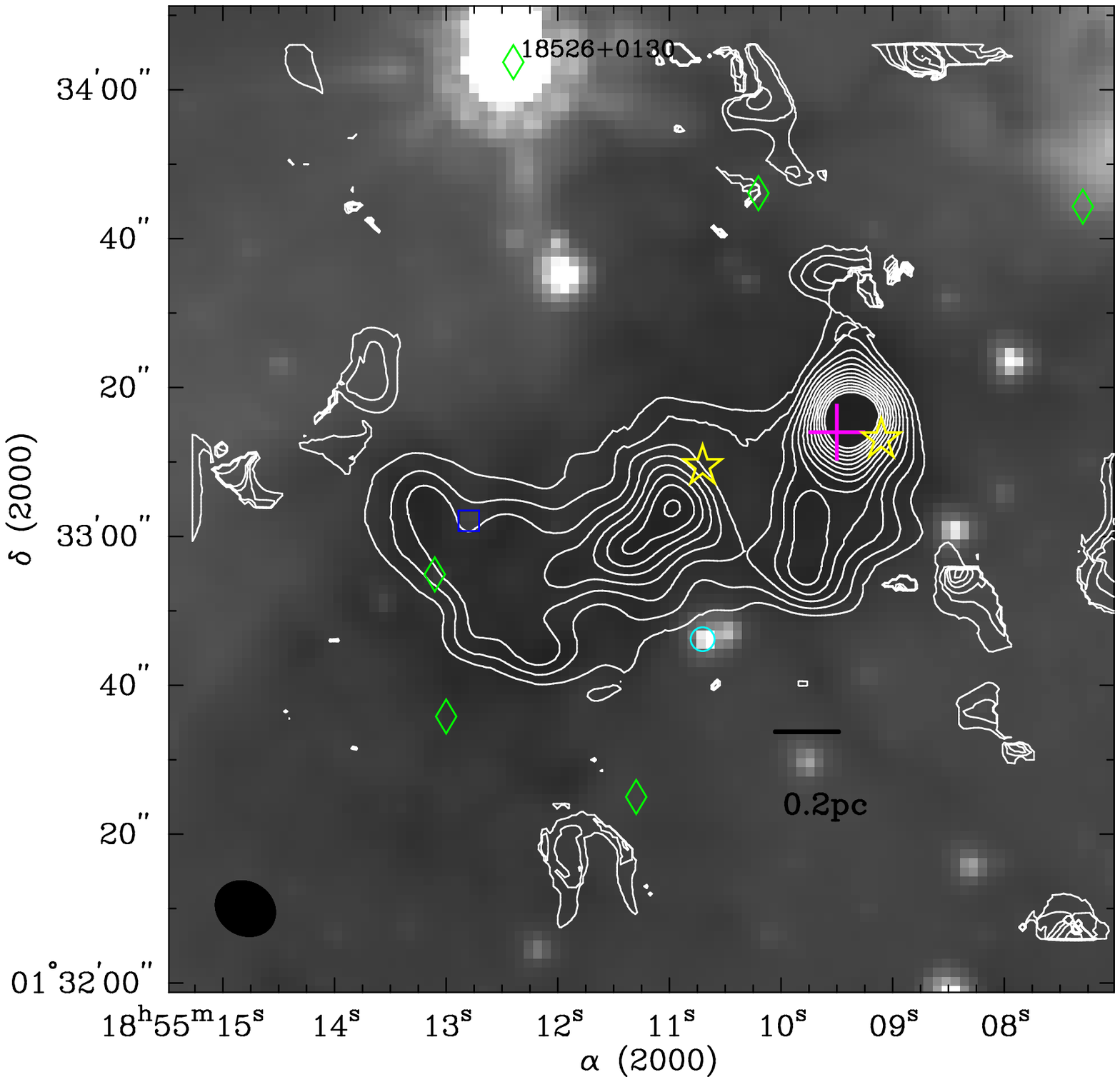,height=5.0cm,angle=270}
}
\caption{NH$_3$ column density contours plotted over the {\em Spitzer} 8$\mu$m image of the IRDC.  The magenta crosses mark the locations of the emission peaks described in Table 2. The other symbols signify the locations of YSOs as follows: Class I (blue square), Class II (green diamond), transition disk objects (cyan circles), embedded objects (red diamonds), and unclassified point sources appearing only at 24~$\mu$m (star symbols).
Top Left: G005.85$-$0.23 N(NH$_3$) contour levels:  2 to 10 $\times$ 10$^{15}$ by 1 $\times$ 10$^{15}$ cm$^{-2}$.
Top Middle: G009.28$-$0.15 N(NH$_3$) contour levels: 3 to 17 $\times$ 10$^{15}$ by 2 $\times$ 10$^{15}$ cm$^{-2}$.
Top Right: G009.86$-$0.04 N(NH$_3$) contour levels: 2 to 6 $\times$ 10$^{15}$ by 1 $\times$ 10$^{15}$ cm$^{-2}$.
Bottom Left: G023.37$-$0.29 N(NH$_3$) contour levels:  3 to 18 $\times$ 10$^{15}$ by 5 $\times$ 10$^{15}$ cm$^{-2}$.
Bottom Middle: G024.05$-$0.22 N(NH$_3$) contour levels: 2 to 10 $\times$ 10$^{15}$ by 1 $\times$ 10$^{15}$ cm$^{-2}$.
Bottom Right: G034.74$-$0.12 N(NH$_3$) contour levels: 8 to 30 $\times$ 10$^{15}$ by 2 $\times$ 10$^{15}$ cm$^{-2}$.
} 
\label{fig:column_contours}
\end{figure*}

\begin{figure*}
\hbox{
\vspace{1.0cm}
\psfig{figure=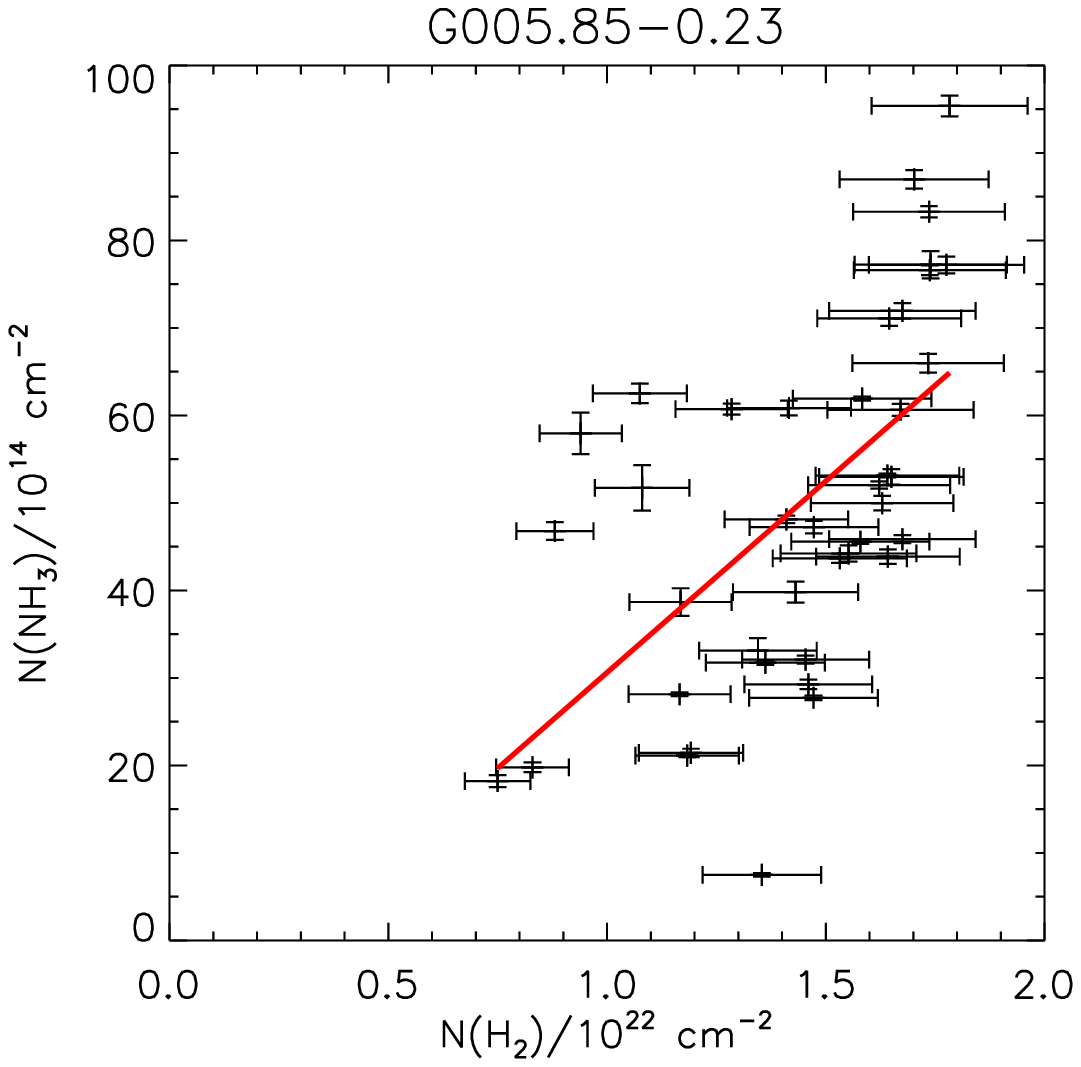,height=6.0cm}
\hspace{-3cm}
\psfig{figure=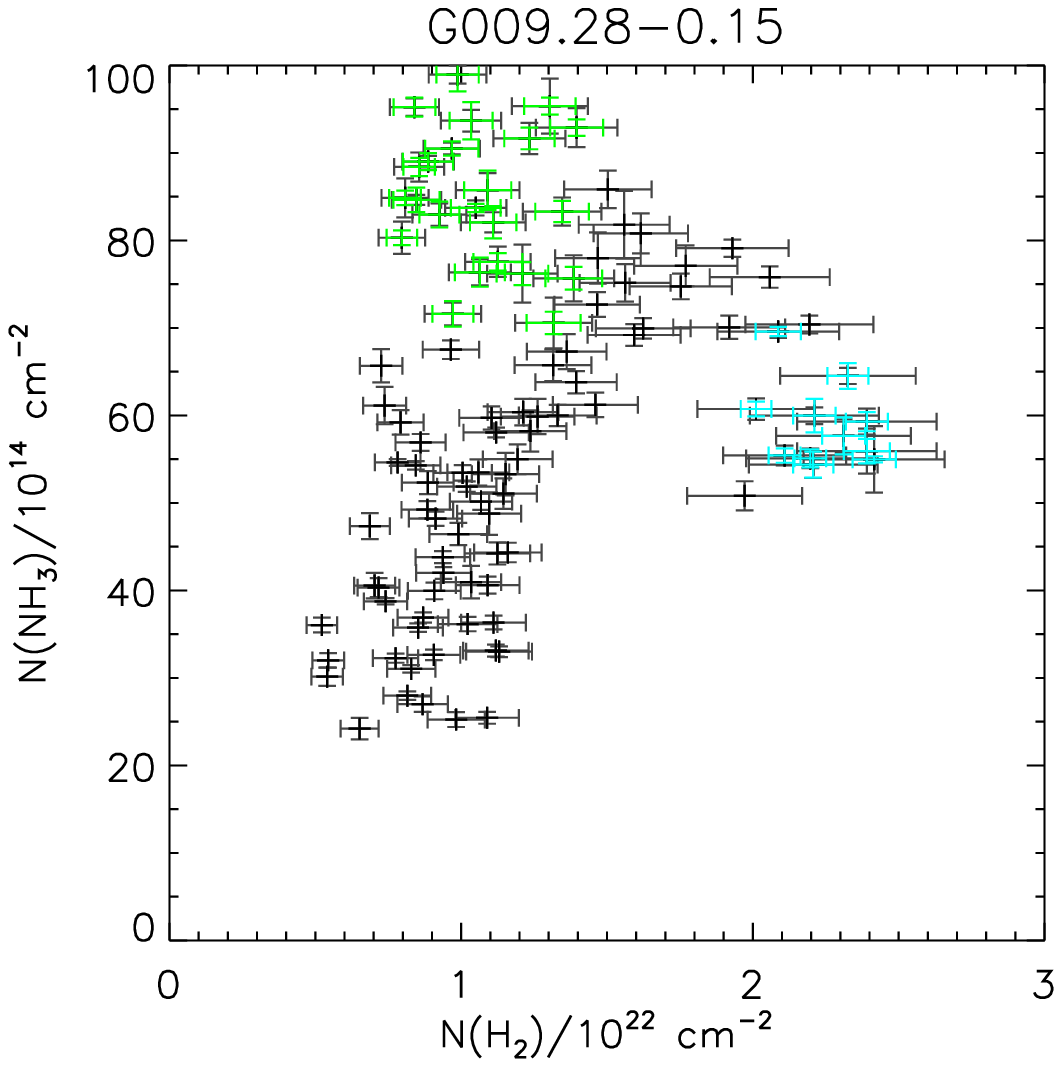,height=6.0cm}
\hspace{-3cm}
\psfig{figure=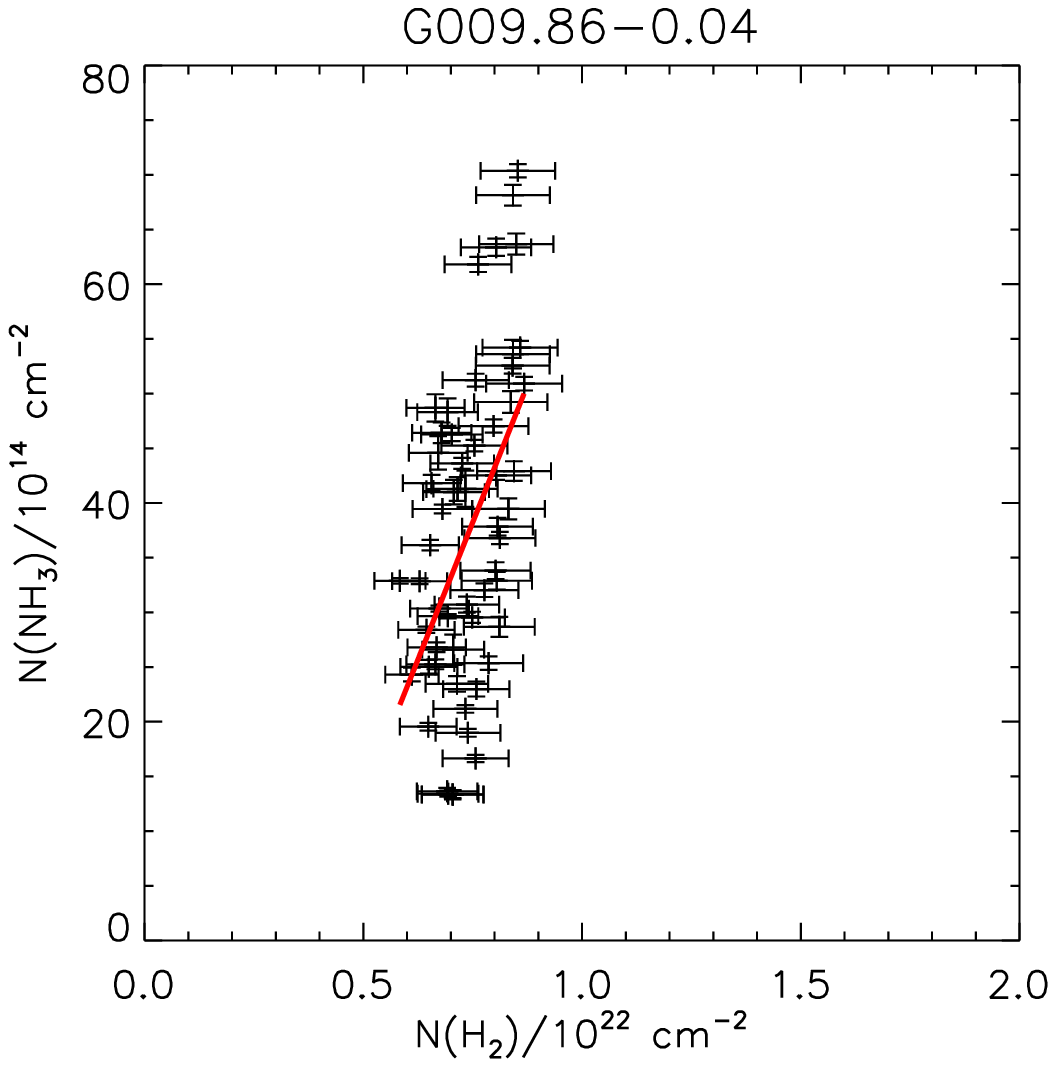,height=6.0cm}
}
\vspace{-1.0cm}
\hbox{
\vspace{1.0cm}
\psfig{figure=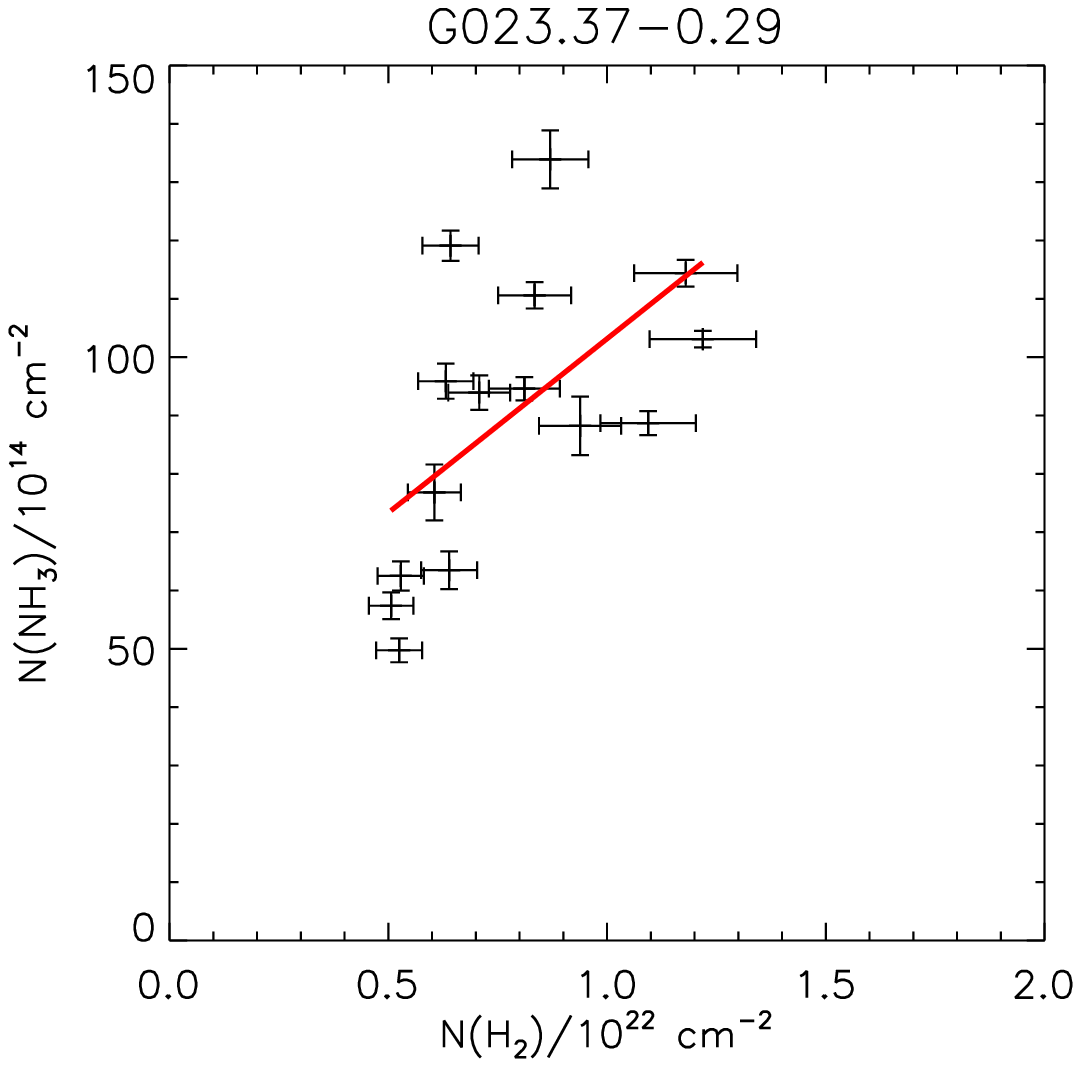,height=6.0cm}
\hspace{-3cm}
\psfig{figure=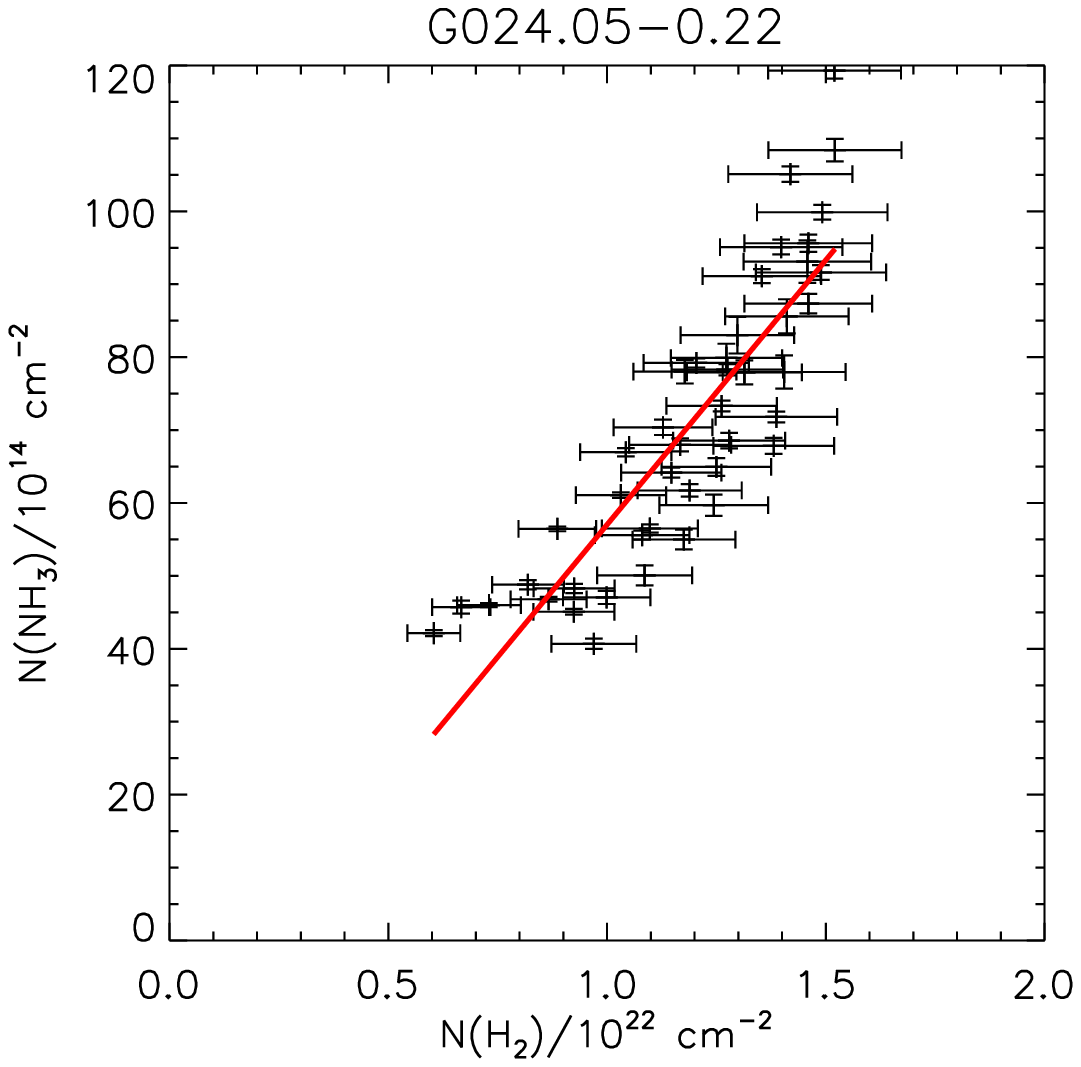,height=6.0cm}
\hspace{-3cm}
\psfig{figure=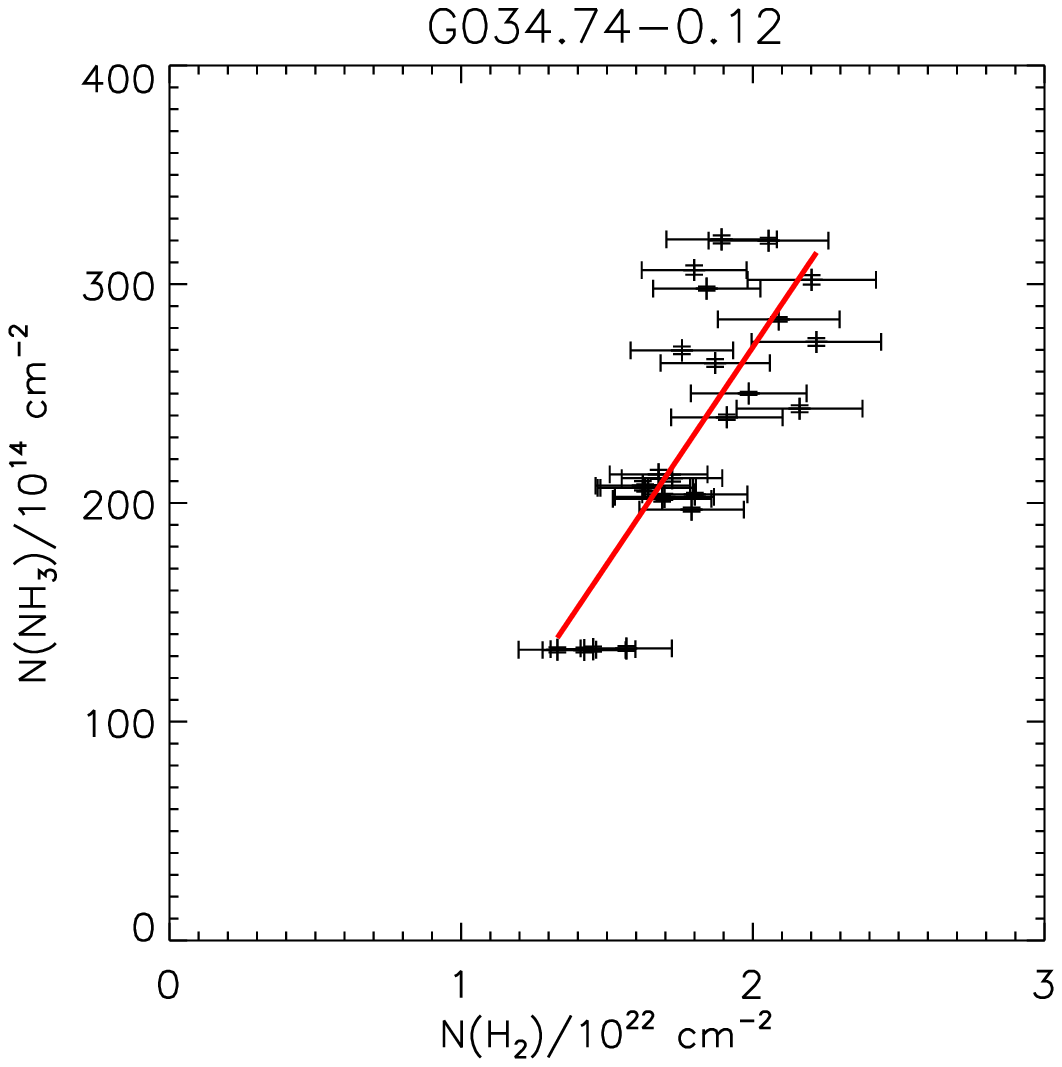,height=6.0cm}
}
\caption{NH$_3$ column density versus the H$_2$ column density (10\% error bars shown). The red line shows the trend of the linear fit of the abundance, $\chi_{NH_3}$ = N(NH$_3$)/N(H$_2$). The average abundance in this sample is 8.1 $\times$ 10$^{-7}$. The fits were only performed in unsaturated regions of moderate total optical depth in the (1,1) line ($\tau(1,1,m) < 5$). For G009.28$-$0.15, a linear fit did not converge. We show in green and cyan the values corresponding to the regions circled in green and cyan in Figure~\ref{fig:column_contours}. These regions appear to have anomalously high or low NH$_3$ emission (respectively), and do not keep to the one-to-one relation the other IRDCs exhibit. G023.37$-$0.29 is only fit for the un-saturated region around the cloud edges.}
\label{fig:column_plots}
\end{figure*}

\section{Results}

\begin{table*}
\begin{center}
\caption{Peak properties, abundances, and mean temperature of IRDCs \label{tab:abund}}
\begin{tabular}{lccccccc}
\hline
 & \multicolumn{5}{c}{Peak NH$_3$ $\int T dv$ position} &   \\
& \cline{1-5} & & & & &   \\
IRDC & RA & DEC & v & dv & $\tau_{m}(1,1)$ & $\chi_{NH_3}$ & $<T_k>$   \\
& (J2000) & (J2000) & (km s$^{-1}$) & (km s$^{-1}$) & & (10$^{-7}$) & (K)   \\
\hline
G005.85$-$0.23 & 17:59:51.4 & -24:01:10 & 17.4 & 1.4 & 4.9 & 4.4 & 11.0$\pm$1.0  \\
G009.28$-$0.15 & 18:06:50.8 & -21:00:25 & 41.0 & 1.5 & 4.0 & \nodata\tablenotemark{a}& 11.2$\pm$1.3  \\
G009.86$-$0.04 & 18:07:35.1 & -20:26:09 & 17.8 & 1.4 & 3.3 & 5.2 & 10.3$\pm$1.1  \\
G023.37$-$0.29 & 18:34:54.1 & -08:38:21 & 78.5 & 3.7\tablenotemark{b} & \nodata\tablenotemark{b} & 5.4 & 8.8$\pm$3.3  \\
G024.05$-$0.22 & 18:35:54.4 & -07:59:51 & 81.7 & 2.0 & 2.6 & 6.7 & 10.1$\pm$1.4  \\
G034.74$-$0.12 & 18:55:09.5 & +01:33:14 & 78.0 & 2.3 & 6.1 & 19. & 8.5$\pm$0.8  \\
\hline
\end{tabular}
\end{center}
\tablenotetext{a}{ Linear fit did not converge.}
\tablenotetext{b}{ Line is saturated and optically thick at emission peak. Abundance fit in cloud edges.}
\end{table*}

The combined data were made in to spectral line maps in both the (1,1) and (2,2) transitions in a single VLA field of view ($\sim$2$'$ at 1.3\, cm) for each IRDC. Spectra at each position were fit with a customized gaussian minimization routine, and the peak flux, centroid velocity, and linewidth were measured. As the VLA backend configuration did not fully cover the (1,1) spectral signature, when estimating the integrated intensity, we assume the intensity ratio described in \citet{HoTownes}, where the outermost satellite lines account for 0.22 of the intensity compared to that of the main line. The total NH$_3$ (1,1) integrated intensity does not factor into the calculation of the column density or temperature discussed in the present work. In Table~\ref{tab:abund} we list the position of the NH$_3$(1,1) emission peak, the velocity and linewidth there, and the total ammonia abundance and average temperature over the whole mapped region.

\subsection{Ammonia as a tracer of high density gas in IRDCs}

Ammonia has long been used as a reliable tracer of high density gas in studies of low-mass cores \citep{Jijina1999, Rosolowsky_perseus, Friesen2009} and high-mass cores \citep{Harju:1993, Molinari1996,Wu_ammonia}, where other molecules, such as CO, are depleted \citep{bl97}.  The initial studies of ammonia in IRDCs \citep{Pillai_ammonia, Wang_ammonia} show that ammonia is also a good tracer in these environments that have the potential to form clusters containing massive stars.  The ammonia column density can be found through analysis of the (1,1) hyperfine structure, the derivation of which can be found in Appendix A.1.

In Figure \ref{fig:column_contours}, we show ammonia column density contours plotted over the 8~$\mu$m {\em Spitzer} image for each IRDC.  The resolution element is shown in the lower-left corner of each panel. In Figure \ref{fig:column_plots}, we plot the column density of NH$_3$ as a function of H$_2$ column density for each IRDC.  The H$_2$ column density is calculated from the absorption in the {\em Spitzer} IRAC 8~$\mu$m image\footnote{The H$_2$ column densities fall below the 10$^{23}$~cm$^{-2}$ ``saturation limit'' reported by \citet{PerettoFuller2009}, indicating that it is a faithful tracer of H$_2$ in these IRDCs.} \citep[see][]{Ragan_spitzer}. For these analyses, the H$_2$ column density map was convolved to match the VLA resolution for each IRDC, thereby lowering the inferred column \citep[see][]{Vasyunina2009}. In Figures \ref{fig:column_plots}, we do not plot data from regions where the NH$_3$ line opacity exceeds 5, as it is not used in our abundance fits (see below). 

In most cases we find a one-to-one relationship between these two quantities.  The slope of the linear relation is an estimate of the ammonia abundance, $\chi_{NH_3}$. Our fits only include regions in the cloud that have low line optical depths ($\tau_m (1,1)~\lesim 5$). Individual abundance values are given in Table~\ref{tab:abund}, except G009.28-0.15 for which no linear fit was possible. In this case, although most points (in black) show a one-to-one relation similar to the other IRDCs. We investigated where the apparently anomalous emission originated and found that ``high'' values (in green) are confined to a southern region and the ``low'' values are computed for the northern tip. Apart from being contiguous, these datapoints are not otherwise distinct from the rest of the cloud (i.e. not an optical depth effect) or the rest of the sample, therefore we do not discard them for the sake of the desired fit. 
Our restriction on the line opacity most heavily affects the data fit for G023.37-0.29, which excludes all of the central region where the NH$_3$ (1,1) lines are saturated. \citet{Ragan_spitzer} also report the highest values of $\tau_8$ which suggests that N(H$_2$) in the central part of G023.37-0.29 is also an underestimate. Thus the abundance here is calculated for the outer edges of the cloud where both tracers are least hampered by optical depth effects.

The IRDCs G005.85-0.22, G009.86-0.04, G024.05-0.22, G034.74-0.12, and the outer regions of G023.37-0.29 show good correlation between the column densities of H$_2$ and ammonia.  The average of these five IRDC abundances is 8.1 $\times$ 10$^{-7}$, plotted in red on each panel.  IRDCs with no evidence of embedded star formation activity show the tightest relation between ammonia and H$_2$ column densities. 

Our abundance value is higher than 4 $\times$ 10$^{-8}$ found by \citet{Pillai_ammonia} in a single-dish study of ammonia in IRDCs. Though we report similar H$_2$ column densities, we probe higher NH$_3$ column densities, namely because our study has effective beams at least five times smaller than single-dish studies. In a single-dish study, the compact emission is beam-diluted which would result in lower NH$_3$ intensity measurements than what we find here with the VLA. 

\subsection{Temperature maps}

The temperature in cold molecular gas is attainable with knowledge of the optical depth of the line and the ratio of NH$_3$ (1,1) and (2,2) central line intensities.  The derivation of the kinetic temperature is derived in Appendix A.2.

\begin{figure*}
\begin{center}
\includegraphics[scale=0.7,angle=270]{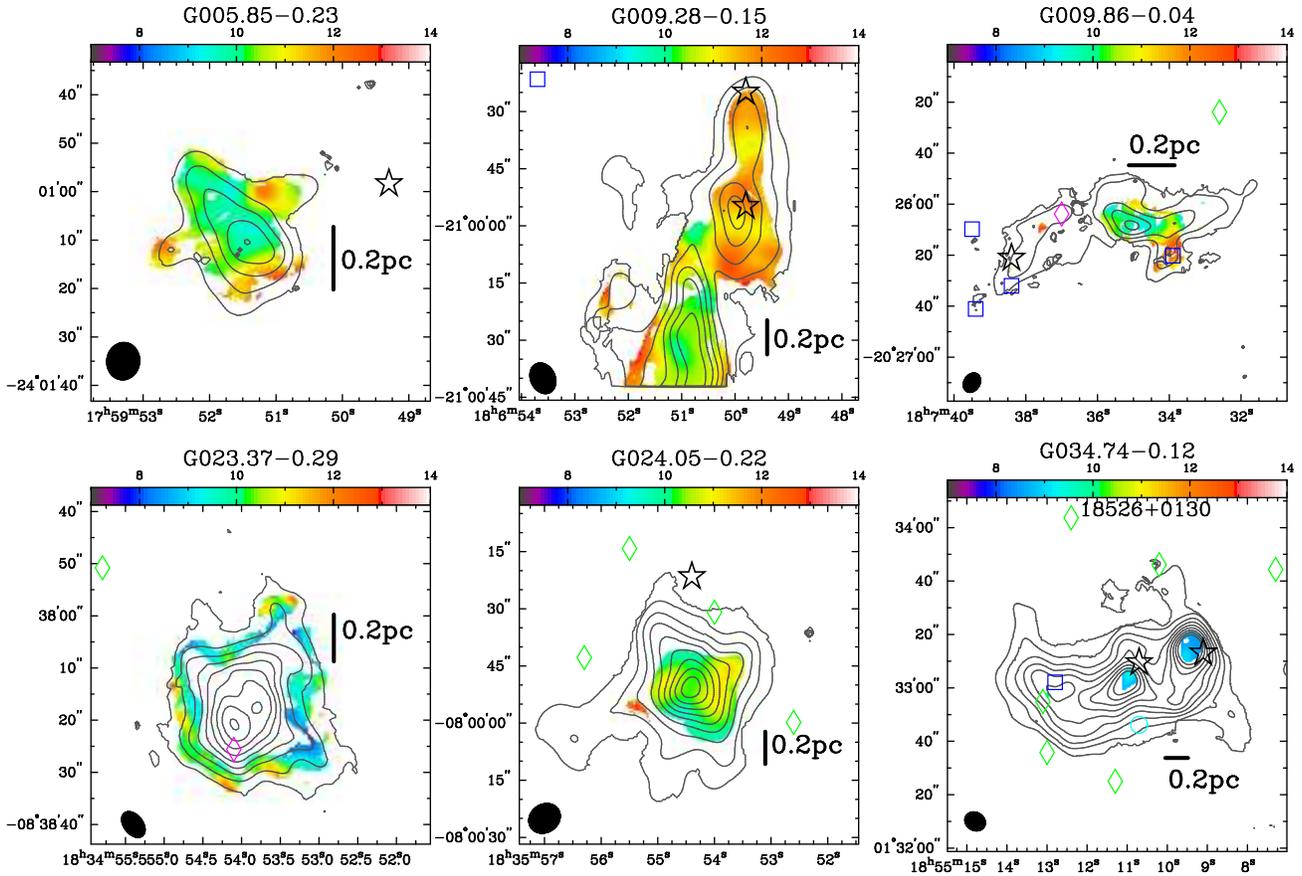}
\end{center}
\caption{Kinetic Temperature map of IRDCs.  Overplotted are contours of the NH$_3$ (1,1) integrated intensity, beginning at 0.2 Jy beam$^{-1}$ km s$^{-1}$ and increasing in steps of 0.1 Jy beam$^{-1}$ km s$^{-1}$.   The 0.2~pc linear scale is indicated on each panel.  Star symbols indicate the location of 24~$\mu$m point sources.  Other symbols are as follows: red diamonds are embedded objects, green diamonds are Class II objects, and blue squares are Class I objects, all as classified in \citet{Ragan_spitzer}.  Only one known IRAS source is present in the fields of the sample: IRAS 18526+0130 which is labeled in the lower right panel near G034.74-0.12.}
\label{fig:tkinmaps}
\end{figure*}

\begin{figure*}
\hbox{
\hspace{-2cm}
\vspace{1.0cm}
\psfig{figure=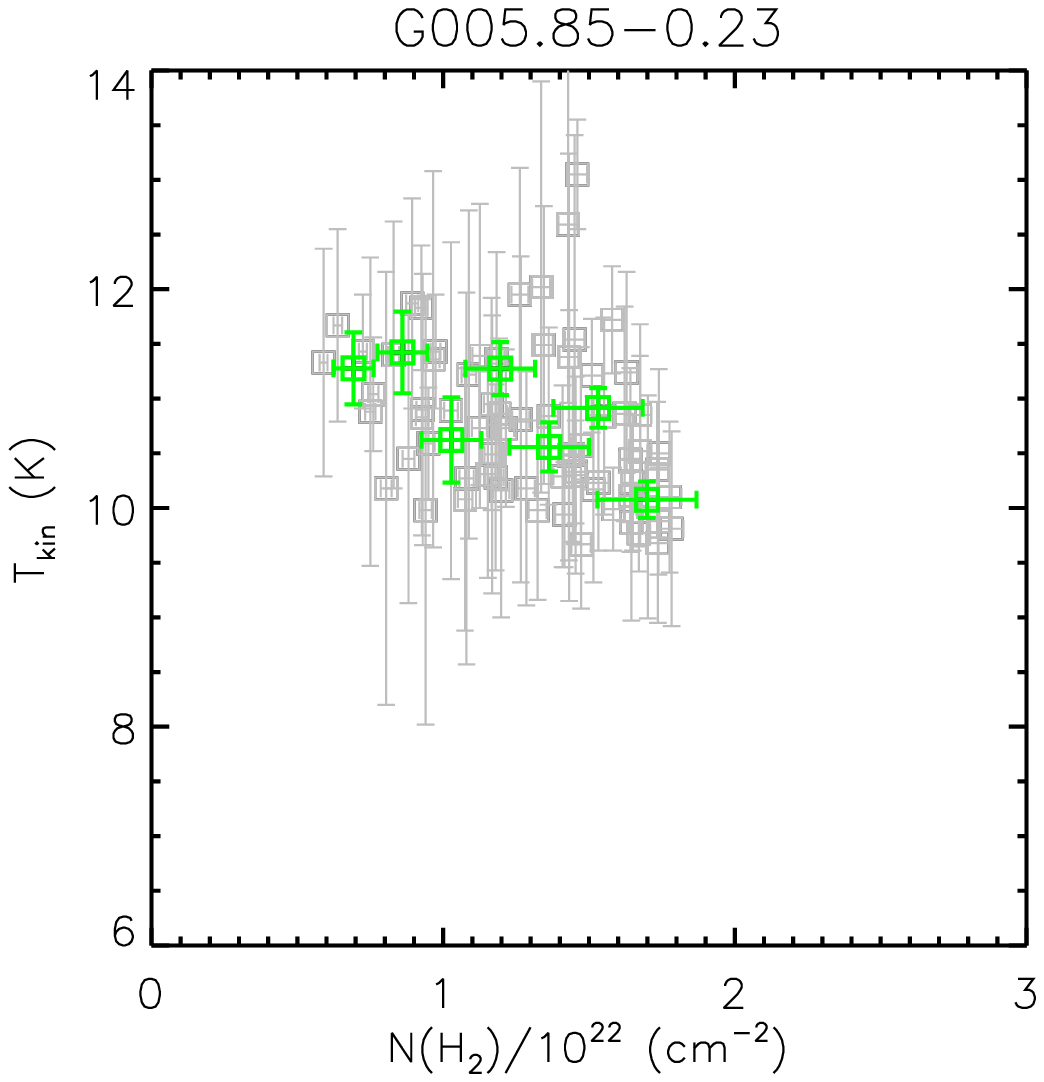,height=6.0cm}
\hspace{-3cm}
\psfig{figure=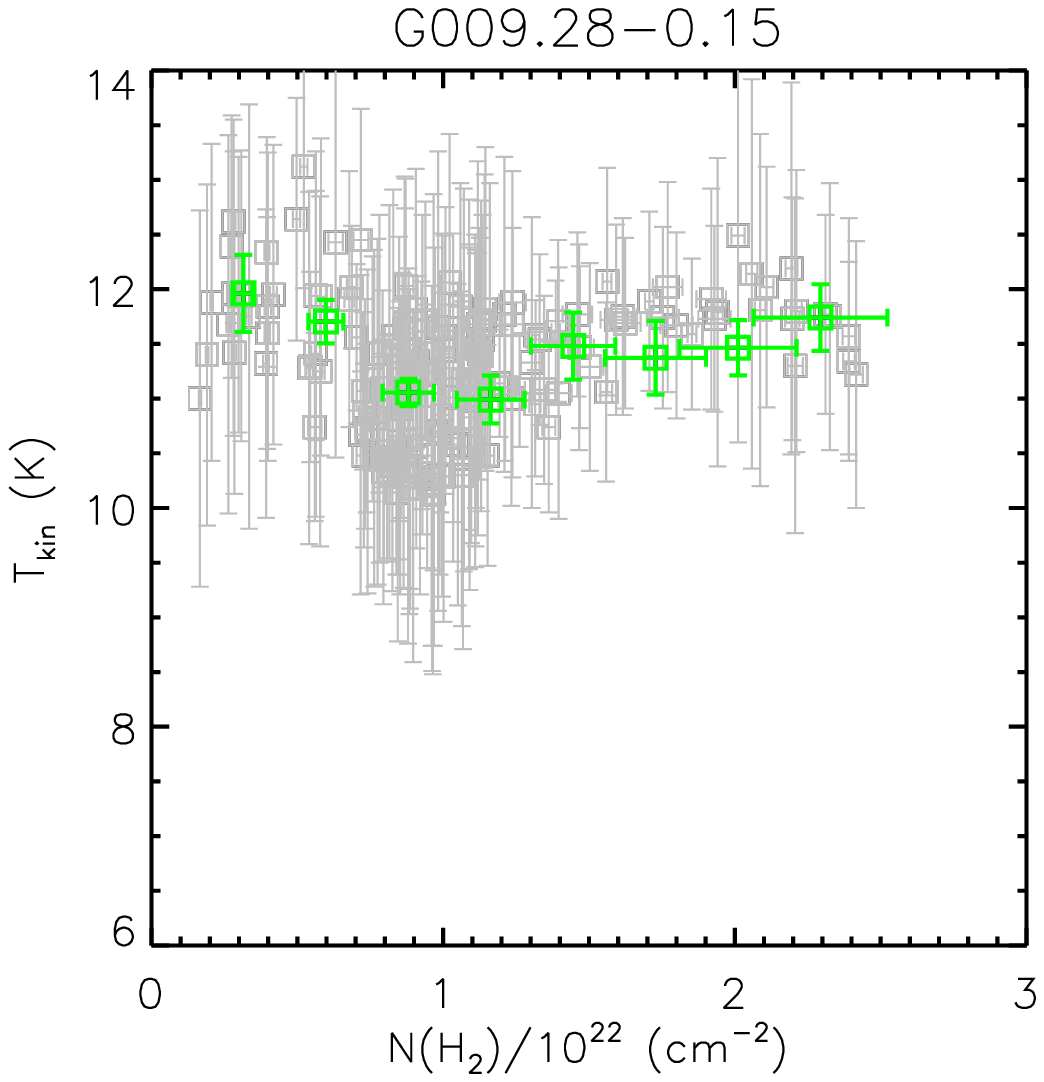,height=6.0cm}
\hspace{-3cm}
\psfig{figure=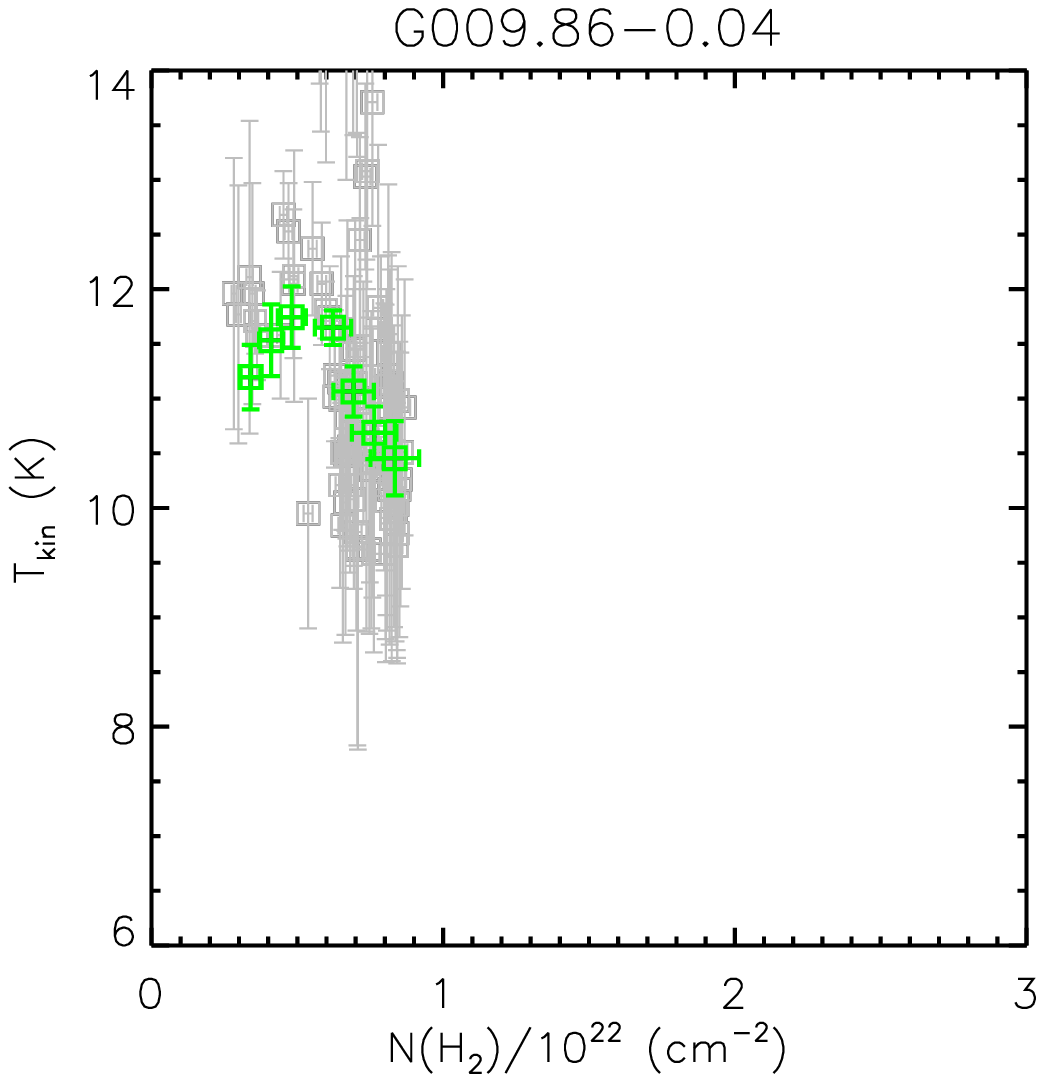,height=6.0cm}
}
\vspace{-1.0cm}
\hbox{
\hspace{-2cm}
\vspace{1.0cm}
\psfig{figure=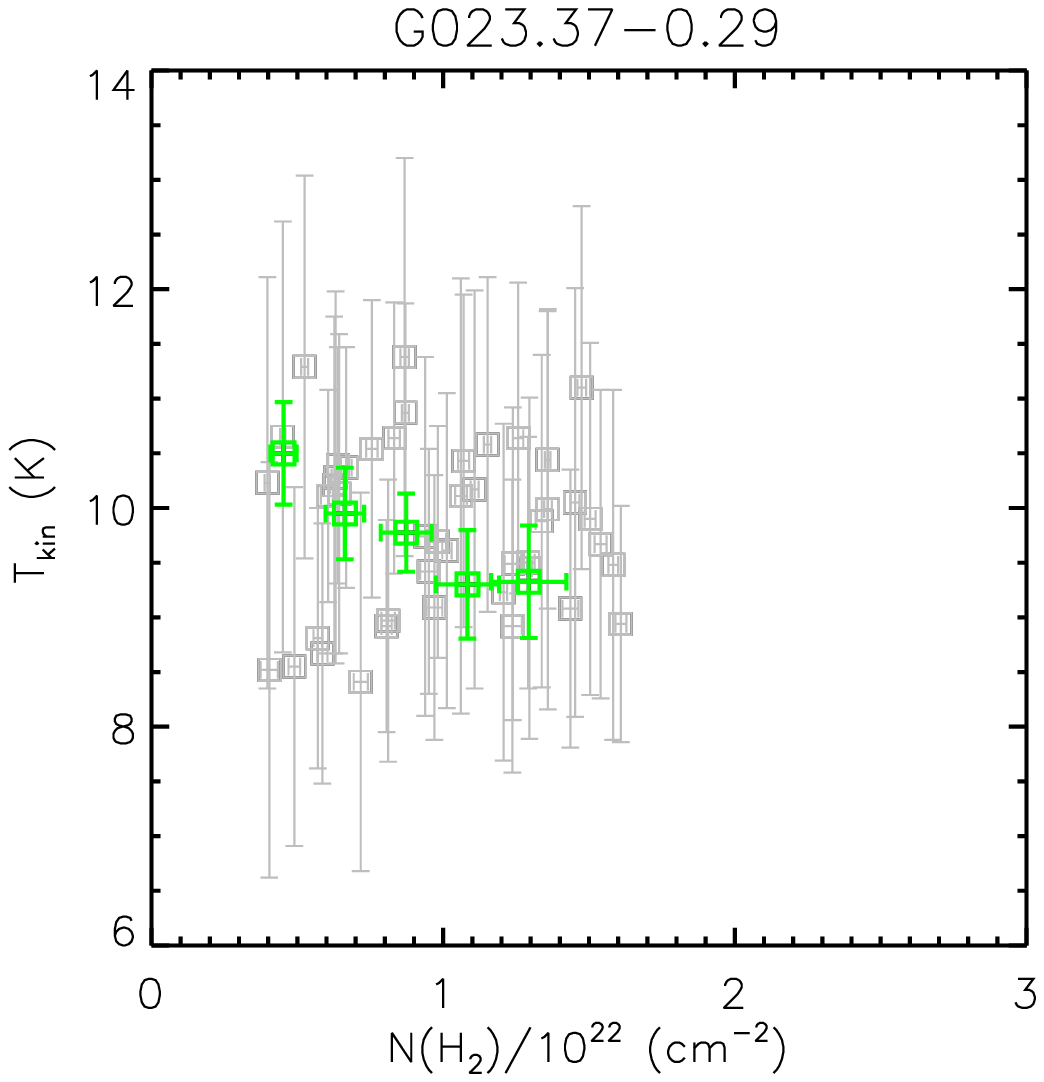,height=6.0cm}
\hspace{-3cm}
\psfig{figure=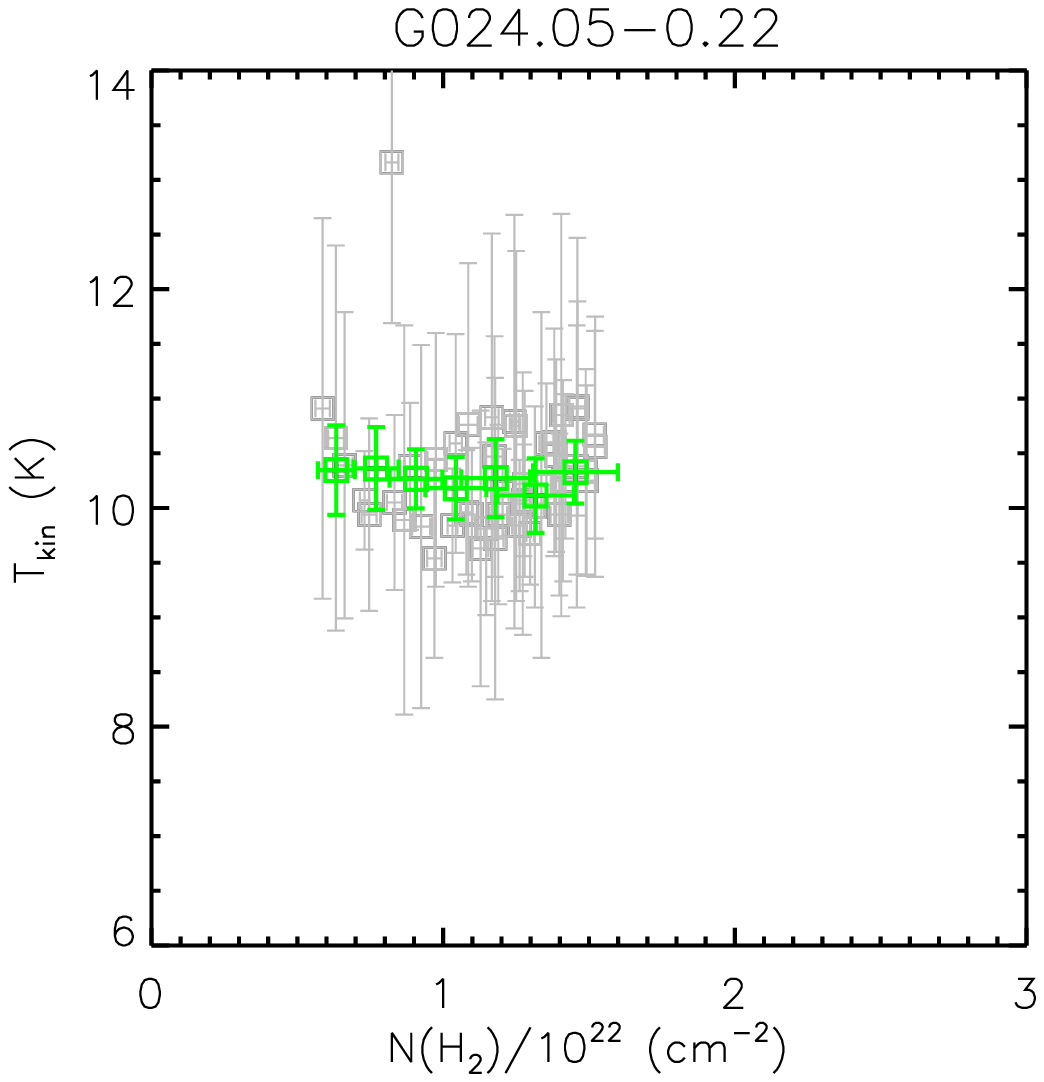,height=6.0cm}
\hspace{-3cm}
\psfig{figure=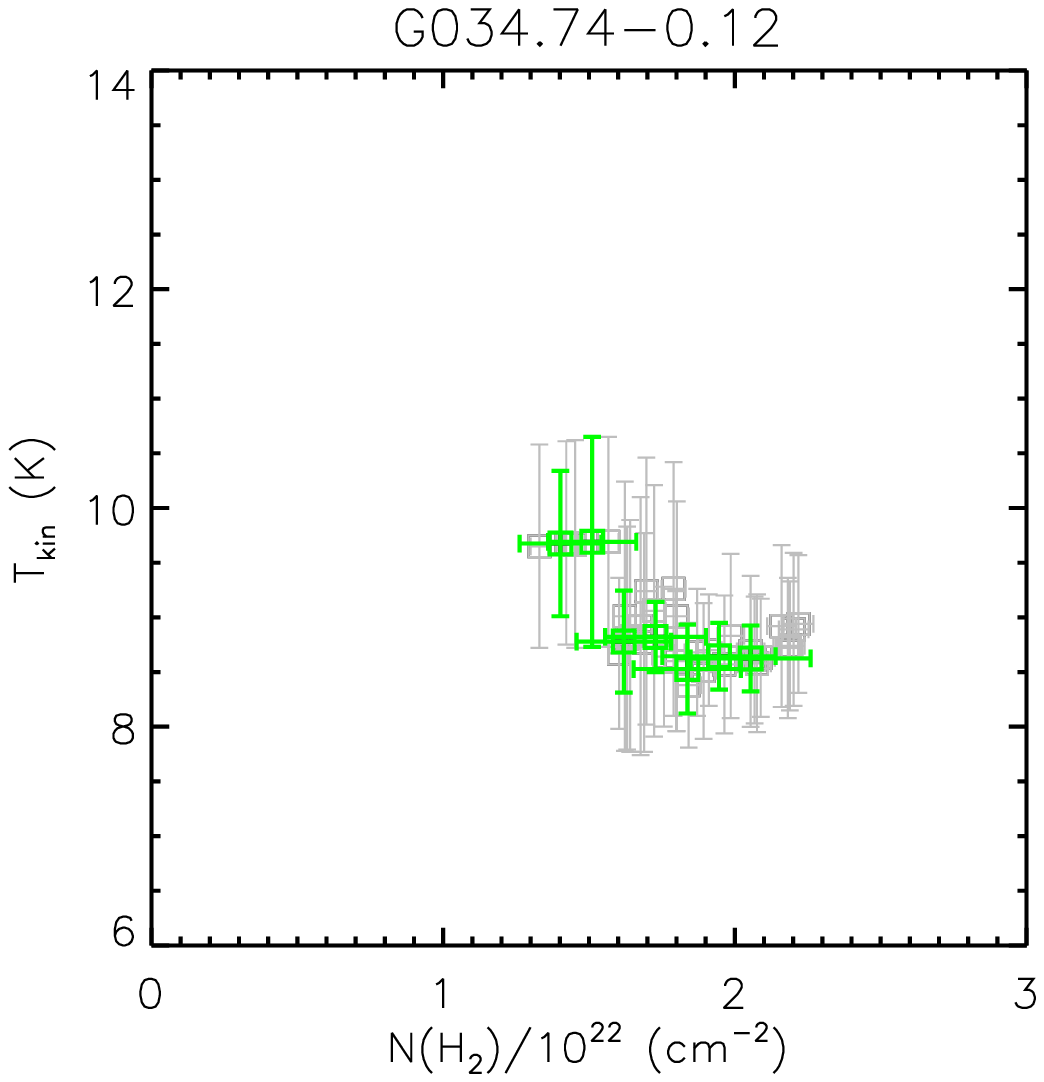,height=6.0cm}
}
\caption{Kinetic temperature (T$_{kin}$) as a function of H$_2$ column density (N(H$_2$)) calculated from 8~$\mu$m absorption.  The green points indicate the weighted average trend of the data.
}
\label{fig:tkintau8}
\end{figure*}

This sample contains objects both with and without signatures of star formation.  In Figure~\ref{fig:tkinmaps}, we present the kinetic temperature maps in all six IRDCs with contours showing the intensity of NH$_3$ (1,1) and symbols indicating the locations of young stars.  The class II sources (the most evolved YSOs displayed) are the most widely distributed population and do not follow the dense gas traced by ammonia.  As such, they are unlikely to have a strong impact of the gas compared to the embedded protostars.
The temperature is only calculated for positions with detections three times the rms noise.  Since the (2,2) emission is typically much weaker than the (1,1) emission, the temperature map boundaries are limited by the (2,2) signal-to-noise.  Kinetic temperatures range from 8 to 13~K.  
In Figure~\ref{fig:tkintau8}, we present the kinetic temperature as a function of the molecular hydrogen column density (N(H$_2$)) derived from 8~$\mu$m absorption\footnote{Were IRDCs round, N(H$_2$) could be used as proxy for radius, thus providing a radial temperature profile.  However, these objects are not spherical and tend to be filamentary in nature, so regions of high optical depth are those most akin to the ``center'' of a dark core or clump, and low optical depth regions are those nearer to edges or near radiation sources such as embedded stars.}\citep{Ragan_spitzer}. 
Table~\ref{tab:abund} lists the average temperature associated with each IRDC.  

IRDCs G005.85-0.23 and G024.05-0.22 lack any 24~$\mu$m point sources coincident with the gas traced by NH$_3$.  We infer that these IRDCs contain no embedded star formation, though there are stars projected in the vicinity (shown on Figure~\ref{fig:tkinmaps}).  In both cases, the NH$_3$ intensity peaks at the same position of both the 8 and 24~$\mu$m absorption peaks.  The average temperature profiles (Figure~\ref{fig:tkintau8}) in these two starless IRDCs are the flattest of the entire sample ($<$T$_{kin}$ $>$ varies by $\sim$1~K in both cases), which is consistent with these objects being least influenced by radiation sources.


The four remaining IRDCs in the sample (G009.28-0.15, G009.86-0.04, G023.37-0.29, and G034.74-0.12) coincide with at least one 24~$\mu$m point source, indicating that these objects likely host embedded protostars. The ``northern'' region of G009.28-0.15 (shown in cyan in Figures~\ref{fig:column_contours} and \ref{fig:column_plots}), which coincides with two 24~$\mu$m point sources, appears to have a slightly elevated temperature compared to the ``southern'' region (shown in green in Figures~\ref{fig:column_contours} and \ref{fig:column_plots}), which appears infrared-dark. While these regions are each spatially coherent, the absolute difference in the mean temperature is within the temperature measurement error shown in Figure~\ref{fig:tkintau8}\footnote{In order to minimize the random errors and retain the largest possible spatial expanse of reliable temperature measurements, we binned the spectra to 0.62~km s$^{-1}$ resolution.}.

There are six 24~$\mu$m point sources in G009.86-0.04, five of which are concentrated in the eastern portion of the cloud where it is not possible to calculate the temperature because of very weak (2,2) emission.  The kinetic temperature at the integrated intensity peak is 9~K, and the point source in the to the southwest has an elevated temperature of 13~K.  We see the strongest temperature gradient in this source (see Figure~\ref{fig:tkintau8}), with the lowest temperatures tending toward the highest optical depths.

IRDC G023.37-0.29 exhibits the strongest NH$_3$ emission in the entire sample, and we also estimate the highest abundance in this object.  The lines near the center of this object are saturated, indicating that the emission is very optically thick.  The temperature then is likely not representative of the conditions near to the 24~$\mu$m source near the center of the object, which could explain why we do not see any elevation in kinetic temperature like we do in the other IRDCs in the sample.  It is likely that this object would potentially exhibit strong emission in higher $J, K$ lines and, therefore, be at a higher temperature not probable by the present observations.

In the IRDC G034.74-0.12, emission is widespread in both the (1,1) and (2,2) transitions. We find the highest ammonia column densities in this object ($\sim$ 3$\times$ 10$^{16}$ cm$^{-2}$. This map is also the noisiest, and the (2,2) emission map only passed our 3-sigma detection limit test at the two strong peaks also in (1,1) emission. If the noise requirement for the (2,2) emission is relaxed, the resulting temperature across the source is very uniform, between 8 and 9~K.  Four 24~$\mu$m point sources are detected on the IRDC, and the IRAS point source 18526$+$0130 ($L_{bol} \sim 10^4 \lsun$ \citet{Robitaille2007}) is 30$\arcsec$ north of the NH$_3$ map boundary, corresponding to 0.7~pc at the distance to this IRDC, though no direct influence of the IRAS source is evident in our temperature measurements.

\section{Discussion}

\subsection{Column density structure}

Throughout studies of star formation regions with molecular emission, ammonia has been one of the best tracers of cold dense gas, showing no compelling evidence of depletion up to very high densities, $n >10^5$ cm$^{-3}$ \citep{bl97} .  The same holds true in IRDCs, though in regions of very high H$_2$ column density, the ammonia line can become optically thick.  We find no evidence of depletion of ammonia in IRDCs, thus it can serve as a reliable tracer of kinematics throughout the bulk of the gas which is not yet forming stars. We will, in the following sections, discuss the limitations of the (1,1) and (2,2) lines in probing warm gas in compact regions near embedded protostars.

We note that our estimated NH$_3$ abundances are high when compared to other cold objects.
For example, \citet{taf04} find central abundances on the order of $\sim 2 \times 10^{-8}$ in a few cold
low mass cores in Taurus, while \citet{Harju:1993} find values from $1 - 5 \times 10^{-8}$ in a sample of isolated cores
in Orion and Cepheus.  Interestingly, \citet{taf04} find evidence for an increase in the NH$_3$ abundance
in the denser gas.  They postulate that perhaps greater CO depletion might lead to enhancements of the ammonia
abundance via gas phase reactions.  Our sources likely have greater densities and perhaps that induces greater
formation of ammonia.  However, this question needs to be addressed with more complete modeling spanning
a wide range of densities.

\subsection{IRDC temperatures}

We find that the temperatures in this sample of IRDCs, despite spanning from quiescent to active phases, are largely uniform in the bulk of the cloud, ranging only from 8 to 13~K.  These results are in general agreement with the single-dish measurements of temperature in IRDCs by \citet{Pillai_ammonia}, and also the temperatures found in Perseus \citep{Rosolowsky_perseus} and Ophiuchus \citep{Friesen2009}.  

Ammonia studies of temperature structure in {\it massive} star forming complexes have mainly focused on Orion. \citet{Wiseman1998} map the filamentary region surrounding the Orion-KL massive star forming core complex.  One source within this complex is IRc2, which drives a large bipolar outflow.  The surrounding molecular ridge, OMC-1, is comprised of several finger-like filaments in which they find the kinetic temperature ranges between 15 and 30~K.  In the central core around Orion-KL, they find temperatures near 80~K, and near the edges of the filaments, the temperature is elevated to $\sim$40~K.  The heating is detectable up to 0.3~pc away from the central core, and is attributed to radiation and winds from Orion-KL sources and shock interactions near the core where the highest temperatures are found.  If such luminous sources were present in IRDCs, we would see greater temperature variation. The low, unvarying temperature of the bulk of the IRDC gas indicate that IRDCs are in an earlier evolutionary stage than Orion.

\citet{Li_oriontemp} select quiescent cores far away from the Orion-KL region, a site of active massive star formation.  The authors observe a similar temperature trend as the one presented here: the lowest temperatures are found in the cloud interiors and warmer gas is near the edges.  They attribute this to external heating from the elevated radiation field due to nearby active star formation region in the Trapezium, despite the large amount of extinction surrounding the region probed by ammonia. These quiescent Orion cores can be most favorably compared to the IRDCs G005.85-0.23 and G024.05-0.22 in our sample.  They show no 24~$\mu$m point sources coincident with the dense gas. We see very little heating in these objects, and by comparing the colored region (where there is a NH$_3$ (1,1) {\it and} (2,2) detection) and the contours (NH$_3$ (1,1) intensity), the region where temperature calculation is possible is already heavily embedded in dense gas, and therefore shielded from the interstellar radiation field.

One high-resolution follow-up study of IRDC G028.34$+$0.06 \citet{Wang_ammonia} showed a significant rise in rotation temperature at the sites of 24~$\mu$m point sources.  In this case, the more evolved, warmer region contains a well characterized high-mass protostellar object and IRAS source (L $\sim$ 10$^3 \lsun$) and has a peak rotation temperature of 30~K, much higher than detected in any IRDC in this study.  In the single-dish study of this object, however, the rotation temperatures across the IRDC were found to be between 13 and 16~K.  Thus, \citet{Wang_ammonia} demonstrate that without the high-resolution of an interferometer, these compact features are diluted by the emission from the cold, quiescent bulk of the IRDC.

Our sample of six IRDCs was selected to be at the earliest evolutionary stage, in regions devoid of maser emission, radio emission, and IRAS sources \citep{ragan_msxsurv}.  
Thus, the lower average kinetic temperature ($\sim$11~K) is consistent with this selection, and this sample of IRDCs are at an earlier stage in their evolution compared to G028.34$+$0.06.  Furthermore, we have no instances of such elevated temperatures such as that seen in \citet{Wang_ammonia}, indicating that nowhere in these IRDCs is there massive star formation at such an advanced stage, if it exists at all.  

The absence of temperature structure does not preclude the presence of star formation. Indeed there exists a surface population of young (low-mass) stars in these IRDCs \citep{Ragan_spitzer}, however, these appear to have a minimal impact on the gas traced with these observations. 
In the cases where we detect embedded star formation, the effect on the temperature is localized and does not influence the widespread dense gas in IRDCs
\footnote{IRDCs show high extinction levels, with typical A$_K$ values in the 1 -- 3 range \citep[or A$_{24}$ from 0.5 to 1.5][]{Flaherty2007} for associated YSOs \citep{Ragan_spitzer} and higher still for more deeply embedded sources not detected by {\em Spitzer}. For example, assuming A$_{24}$=1, or A$_K$=2, at 5\, kpc, up to a 100\, $\lsun$ star could fall below our detection limit.}

\subsection{Fragmentation scale constraints}

The temperatures determined in this study can place constraints on how fragmentation proceeds in IRDCs.  The traditional characteristic fragmentation scale (which assumes isothermality, spherical symmetry, and uniform density) is given by the Jeans length:

\begin{equation}
\lambda_{J} = \sqrt{\frac{\pi c_s^2}{G \rho}}
\end{equation}

\noindent where c$_s$ is the isothermal sound speed (0.18~km s$^{-1}$ for 10~K gas).  For an average density of 10$^5$ cm$^{-3}$ ($\rho \sim 4 \times 10^{-19}$ g cm$^{-3}$), the Jeans length is 0.07~pc, which is below the effective physical resolution of our VLA observations (0.1 to 0.2\, pc).  The corresponding Jeans mass (M$_J$ = $\rho \lambda_J^3$) is about 2~$\msun$.  While the detectable star formation activity within these IRDCs is limited to a few objects, in the cases where there are multiple objects of class I or earlier (G009.28-0.15, G009.86-0.04, G034.74-0.12), the average separation between protostars is 0.4~pc, within a factor of 5 of the Jeans length scale. This results is similar to what \citet{Teixeira2006} found in a filamentary star formation region (with a factor of three of $\lambda_J$), which is roughly consistent with thermal fragmentation.  Given that IRDCs are a several kiloparsec distances, hampering our sensitivity to faint stars, our scale is an upper limit.  

In reality, the material in IRDCs is much more clumped and filamentary than a uniform sphere.  An alternative quantification of such a cloud is the critical fragmentation mass-per-unit-length scale in a filament \citep[e.g.][]{Inutsuka1992, Andre_sdp}, which gives 15~$\msun$ pc$^{-1}$ for 10~K gas (not taking into consideration any magnetic field support).  The mass-per-unit-length in these IRDC filaments are typically a few hundred solar masses per parsec, well in excess of the critical value.  This would indicate that these objects are undergoing collapse and actively forming stars.  The dynamical state of this IRDC sample will be analyzed in detail in a forthcoming publication (Ragan et al., in prep.).

\subsection{The high pressure environment of IRDCs}

In \citet{Ragan_spitzer} we discussed the overall mass distribution of IRDCs in comparison to the Orion Molecular Cloud, which is the local example of massive and clustered star formation.   From our absorption study we can estimate a gas mass surface density for IRDCs, $\Sigma_{IRDC}, between 1000 and 5000$ M$_{\odot}$ pc$^{-2}$, while in Orion the mass surface density is $\Sigma_{OMC}\sim 30$  M$_{\odot}$ pc$^{-2}$.  
Thus IRDCs are not distant Orions, but rather something fundamentally different.  Here we examine what role the environment of IRDCs might play in supporting such a large concentration of material where the prevalence of star formation appears relatively low.

\citet{Jackson_galdistr_IRDCs} has examined the spatial distribution of IRDCs which were previously identified with the Molecular Ring that contains over 70\% of the galactic molecular mass. They argue that the distribution is best fit by a model of the Milky Way with two major spiral arms emerging from a central bar.  The first quadrant IRDCs lie preferentially in the Scutum-Centarus arm.  This is in agreement with other results probing the dust structure in the galaxy \citep[e.g.][]{ds01, benjamin05}.   

Theoretical models of gas dynamics in barred galaxies suggest that gas flow produces arms of gas/dust density enhancement that extend from the leading side of the bar  \citep[][]{athan92}.  These enhancements are the result shocks and other dynamical processes which results in the formation of molecular clouds from the atomic medium \citep{cg85, bergin_cform}.  The IRDC surface density corresponds to an average density of $<n_{\rm H_2}> \sim 10^5$ cm$^{-3}$.  Given a typical uniformly low temperature of $\sim$ 10~K this corresponds to a internal thermal pressure of P$_{int}$/$k$ $\sim$ few $\times 10^6$ cm$^{-3}$ K well in excess of that in Orion, where the gas temperature is typically 20~K, or the general ISM where P$_{ISM}/k = 2240$  cm$^{-3}$ K \citep{jt01}.  These high pressures require a strong dynamical interaction to gather ISM gas \citep[c.f.][]{bergin_cform}, enabling gravitational collapse to commence, and fragmentation to proceed. This interaction may be related to gas flow induced by the central Galactic bar.

\section{Conclusions}

Ammonia is an excellent tracer of the dense gas in IRDCs.  To the resolution of our VLA observations, we find a roughly linear relationship between ammonia and molecular hydrogen column density up to N(H$_2$) $\sim$ 3 $\times$ 10$^{22}$~cm$^{-2}$ as traced by mid-infrared extinction.  We approximate the abundance of ammonia in IRDCs to be 8.1 $\times$ 10$^{-7}$.  We see no evidence of ammonia depletion in the bulk of the gas in IRDCs.

Our NH$_3$ (1,1) and (2,2) observations show that IRDCs have uniform temperatures between 8 to 13~K, varying only by 2 to 4~K within a given cloud.  We see no strong evidence for external heating of the gas in IRDCs as probed by ammonia. IRDCs are embedded in large complexes of gas that shield it from the radiation of local stars.  Embedded protostars, appearing as 24~$\mu$m point sources, and nearby young stars do not affect the temperature of the bulk gas in IRDCs.  These IRDCs are drawn from a sample of objects with no known signposts of massive star formation (e.g. masers, IRAS sources), and the temperature structure also supports the absence of such activity.  At the same time, the limitation of the NH$_3$ (1,1) and (2,2) thermometer in terms of excitation compromises the sensitivity to the warm gas that may reside on the smallest scales.  Observations of more highly excited lines would be a useful test of the existence of a warmer, more compact core near the embedded sources.  

The gas in IRDCs is undergoing collapse and fragmentation, and they will ultimately form clusters containing massive stars.  Their physical properties (e.g. temperature, density) are similar to gas in local star forming regions, but their high-pressure environment allows for there to be much more mass concentrated in a small area. IRDCs primarily reside in the Scutum-Centarus arm as has been suggested in the literature and the high concentration of dense high-pressure molecular gas could be due to particular dynamics of galactic spiral arms on the leading edge of a central bar.

\acknowledgements
We thank the anonymous referee for a detailed review of the manuscript and whose comments significantly improved this paper. S.R. is grateful to Claire Chandler, Rachel Friesen, and John Monnier for help with data reduction, and to Henrik Beuther and Paul Clark for useful discussions. This work was supported by the National Science Foundation under Grant 0707777.

\appendix
\section{Derivation of physical quantities}
\subsection{Column Density}
We calculate the NH$_3$ (1,1) column density following \citet{RohlfsWilson}, taking into account both parity states of the (1,1) energy level \citep{Friesen2009}.  

\begin{equation}
N(1,1) = \frac{8\pi \nu_0^2}{c^2} \frac{1}{A(1,1)} \frac{\frac{g_1}{g_2} + e^{-h\nu_0/(kT_{ex})}}{1 - e^{-h\nu_0/(kT_{ex})}}  \int \tau (\nu) d\nu
\end{equation}

\noindent where g$_1$ and g$_2$ are the statistical weights for the upper and lower states of the NH$_3$ (1,1) transition (equal in the case of our two-level approximation), $A(1,1)$ is the Einstein spontaneous emission coefficient, 1.68 $\times 10^{-7} s^{-1}$, and  $\int \tau (\nu) d\nu = \sqrt{2\pi} \sigma_{\nu} \nu_0 \tau_1 / c$ with $\nu_0$ the rest frequency, $\tau_1$ the total line opacity, and $\sigma_{\nu}$ the dispersion \citep{Rosolowsky_perseus}. T$_{ex}$ is the gas excitation temperature, which can be found by fitting the observed spectrum, $T_{a,\nu}$, for a given frequency

\begin{equation}
T_{a,\nu}^{\*}(J,K) = \eta_{MB} \Phi (J(T_{ex}(J,K)) - J(T_{bg}))(1 - exp(-\tau_{\nu}(J,K)))
\end{equation}

\noindent where $\eta_{MB}$ is the main beam efficiency (0.81), $\Phi$ is the filling factor (assumed here to be 1), $J(T)$ is the Rayleigh-Jeans temperature, and $T_{bg}$ is the background temperature, 2.73~K. Finally, the total column density of ammonia is determined via use of the partition function, $Z$:

\begin{equation}
Z = \displaystyle\sum\limits_{J} (2J+1)~S(J)~~exp \left[ \frac{-h[BJ(J+1)\,+\,(C-B)J^2]}{k\,T_k} \right]
\end{equation}

\noindent where B and C are the rotational constants of NH$_3$, 298192 and 186695~MHz respectively\footnote{http://spec.jpl.nasa.gov/}, and $S(J)$ accounts for the nuclear spin statistics of the system.  The total ammonia column density is then 

\begin{equation}
N(NH_3) = N(1,1) \times \frac{Z}{Z_1}
\end{equation}

\noindent in which the $Z$ in the numerator is summed over J=0, 1 and 2, and $Z_1$ in the denominator is evaluated only at J=1.  At the low temperatures of these IRDCs, we assume these metastable states represent a good approximation to the column density of ammonia.

\subsection{Temperature}
The rotation temperature derived from NH$_3$ inversion lines depends on the optical depth and the ratio of the main lines in the $(J,K)$ = (1,1) and (2,2) signatures.  In turn, the optical depth of the NH$_3$ (1,1) main line ($\tau_m$ (1,1)) can be measured by comparing the ratio of the main line intensity to the satellite line intensity, as described in \citet{HoTownes}.

\begin{equation}
\frac{\Delta T_a^{\*}(J,K,m)}{\Delta T_a^{\*}(J,K,s)} = \frac{1 - e^{-\tau(J,K,m)}}{1 - e^{-a\tau(J,K,m)}}
\end{equation}

\noindent where the left hand side is the ratio of the intensities of the main and satellite lines in the NH$_3$ (1,1) signature, and $a$ is the ratio of the intensities compared with the main line intensity.  With measurements of $\Delta T_a^{\*}$ for the main line and the two inner satellite lines (for which $a$ = 0.28), one can solve for $\tau$ numerically.  The total optical depth of the (1,1) line, $\tau_1$ is twice the main line optical depth \citep{Li_oriontemp}.  The rotational temperature ($T_{Rot}$) characterizes the distribution of molecules in these two energy levels, which are separated by 41.5~K in energy, denoted as $T_0$.  

\begin{equation}
T_{Rot} = - T_0 \div ln \left[ \frac{-0.282}{\tau_m (1,1)} ln \left[ 1 - \frac{\Delta T_a^{\*}(2,2,m)}{\Delta T_a^{\*}(1,1,m)} \times (1 - e^{-\tau_m(1,1)}) \right] \right]
\end{equation}

\noindent Since there are no radiative transitions between the (2,2) and (1,1) levels, their population ratio, represented by T$_{Rot}$, probes the importance of collisions, and thus can be used to estimate T$_{k}$.  This assumption is robust in environments such as pre-stellar cores and IRDCs, where temperatures are typically 20~K and lower \citep{Walmsley1983}.  At such temperatures, only the (1,1), (2,2), and (2,1) states are important, thus \citet{Swift2005} showed that the relationship between the kinetic and rotational temperatures is as follows:

\begin{equation}
T_{Rot} = T_k \left[ 1 + \frac{T_k}{T_0} ln \left( 1 + 0.6 e^{-15.7/T_k} \right) \right]^{-1}
\end{equation}

\noindent \citet{taf04} perform Monte Carlo models including the six lowest levels of para-NH$_3$, varying the temperature between 5 and 20~K and self-consistently calculating the (1,1), (2,2), and (2,1) level populations. They find that the approximation

\begin{equation}
T_k = \frac{T_{Rot}} {1 - \frac{T_{Rot}}{T_0}  ln \left[ 1 + 1.1 e^{-15.7/T_{Rot}} \right]}.  
\end{equation}

\noindent is accurate to 5\% or better in the 5 to 20~K temperature range. Recent work by \citet{Maret_ammonia2009} shows that this calibration of the ammonia thermometer holds with this precision taking into account the most up-to-date collision rates in the literature.  Thus, the two-level approximation for the ammonia thermometer described above is robust.

\end{document}